\documentclass[letterpaper]{article}

\usepackage[T1]{fontenc}
\usepackage{geometry}
\usepackage{setspace}

\usepackage[articletitle=true,maxauthors=20,etalmode=truncate]{achemso}

\usepackage{graphicx}
\usepackage{float}
\usepackage[version=4]{mhchem}
\usepackage{amsmath}
\usepackage{amssymb}
\usepackage{booktabs}
\usepackage{array}
\usepackage{textcomp}
\usepackage{authblk}

\newcommand*{\degC}{\ensuremath{^\circ}C}
\newcommand*{\volpc}{vol\,\%}

\author[1,2]{Niclas Alkazaz}
\author[3]{Robert Zierold}
\author[1,2]{Patrick Huber\textsuperscript{*}}
\affil[1]{Institute for Materials and X-Ray Physics, Hamburg University of Technology, 21073 Hamburg, Germany}
\affil[2]{Centre for X-Ray and Nano Science CXNS, Deutsches Elektronen-Synchrotron DESY, 22607 Hamburg, Germany}
\affil[3]{Center for Hybrid Nanostructures CHyN, Universit{\"a}t Hamburg, Luruper Chaussee 149, 22761 Hamburg, Germany}
\date{*Email: patrick.huber@tuhh.de}

\title{Temperature Flips the Sign of the Hydrogen Response in Nanogranular Palladium Networks}

\begin{document}

\maketitle

\begin{abstract}
In nanogranular palladium networks, hydrogen uptake has two opposing effects on resistance. Absorbed hydrogen scatters electrons within the particles, raising their resistance, while particle expansion narrows the gaps between them and lowers the junction resistance. Such networks are widely studied as resistive hydrogen sensors, and the sign of their net response is usually treated as a fixed property of the architecture. In networks grown by atomic layer deposition on macroporous silicon, we show that changing the operating temperature can reverse this sign at a fixed hydrogen concentration. A model indicates that scattering contributes less as temperature rises, allowing sub-{\aa}ngstr{\"o}m changes in gap width to dominate. Where the two effects cancel, the network absorbs hydrogen yet gives the same reading as in hydrogen-free gas, so a leak could escape detection. Between 25 and 40~\degC{}, this cancellation point is predicted to rise over tenfold, crossing reported leak-alarm setpoints. At 55~\degC{}, the measured response stays negative but passes through a minimum, so a concentration above a setpoint can mimic one below it. Annealing at 200~\degC{} alters the calibration and turns the room-temperature response negative at the lowest concentration tested. Quantitative sensing therefore requires a defined thermal pretreatment and an operating temperature giving distinct signals across the intended concentration range. Response polarity becomes an operating parameter rather than a fixed property. More broadly, any nanoparticle network whose particles deform on sorption may show the same temperature-dependent competition between transport within and between particles, with implications for sensors and switches beyond palladium--hydrogen.

\end{abstract}

\section*{Keywords}
hydrogen sensing, palladium nanoparticles, atomic layer deposition, macroporous silicon, sign reversal

\section{} %

Hydrogen sensors are needed wherever hydrogen is produced, stored, or converted, from leak detection in distribution infrastructure to process monitoring in chemical reactors~\cite{Moller2017HydrogenSustainableCarrier,Qanbar2024HydrogenLeakDetection,hubert2011review}. Resistive hydrogen sensors translate gas concentration into a change in electrical resistance. Because this change also depends on operating temperature and device history, reliable operation requires a calibration that accounts for both. In leak detection, the sensor must respond unambiguously well below the lower flammability limit of hydrogen in air, which is approximately 4~\volpc{} at room temperature~\cite{coward1952flammability}. Reported alarm setpoints for area monitors near indoor hydrogen dispensers are 0.4 and 1.0~\volpc{} \ce{H2}, corresponding to 10 and 25~\% of that limit~\cite{buttner2012sensorworkshop}. In chemical processing, self-adapting reactors that adjust their operation to fluctuating renewable feedstocks depend on such local measurements~\cite{hassenstein2026smart,santra2026hydrogen}.

Palladium is widely used for electrical hydrogen detection near room temperature~\cite{hubert2011review,Koo2020ChemiresistiveHydrogenSensors,darmadi2020review,luo_chemiresistive_2023}. Molecular \ce{H2} dissociates on the Pd surface, and atomic hydrogen occupies interstitial sites to form PdH$_x$. Dissolved hydrogen scatters conduction electrons and increases the resistivity of Pd~\cite{flanagan1991palladium,otterson1969absorption}. Hydrogen uptake also increases the lattice parameter by approximately 6~\% per unit H/Pd~\cite{flanagan1991palladium}. In a continuous Pd film, the increase in resistivity generally dominates the electrical response. In a granular film, lattice expansion can bring neighboring particles closer together, close critical gaps, and open conductive pathways, thereby lowering the network resistance~\cite{favier2001mesowire,xu2005ultrathin,kiefer_transition_2010,lee2011nanogap}. In nanoparticle arrays and films near the percolation threshold, both mechanisms operate, and the measured response reflects their balance~\cite{ramanathan_crossover_2010,xie_response_2012,Das2020,Gupta2014,kim_high-resolution_2019}. Reported Pd architectures are accordingly classified by the sign of their response (Table~\ref{tab:benchmark}). For networks near the percolation threshold, this sign is generally attributed to morphology.

The balance between these contributions is sensitive to changes in particle dimensions. A particle 30~nm in diameter expands by only 0.2~{\AA} at a hydrogen loading of $x=0.01$, yet the tunneling conductance between neighboring particles depends exponentially on the gap width~\cite{simmons1963tunneling}. Sub-{\aa}ngstr{\"o}m expansion can therefore switch a critical contact, whereas the same loading changes the lattice resistivity by only approximately one percent~\cite{otterson1969absorption,toth2005electrolytic}. Hydrogen storage in nanoparticles also differs from that in bulk Pd~\cite{yamauchi2008nanosize}. Surface and subsurface sites can bind hydrogen more strongly than bulk interstitial sites~\cite{pundt2006hydrogen,darmadi2020review}. In the dilute $\alpha$ phase of nanocrystalline Pd, hydrogen also segregates to grain boundaries~\cite{kirchheim1988trapping,lemier2007segregation}. Particle size and mechanical constraints narrow the miscibility gap between the dilute $\alpha$ phase and the hydride $\beta$ phase~\cite{griessen2016thermodynamics,syrenova2015hysteresis,baldi2014insitu,dyck_phase_2024}. These size and interface effects also change how hydrogen uptake varies with temperature. Whether changing the operating temperature can shift the balance between electron scattering and junction conduction far enough to reverse the response of the same network remains an open question.

Such a temperature-dependent polarity would undermine any calibration that assumes a fixed sign. At the null point, the concentration at which the two contributions cancel, hydrogen produces no net resistance change, and the sensor gives the same reading as in hydrogen-free gas. If temperature moves the null point toward an alarm setpoint, a leak near that concentration can go unreported. Where the calibration curve is nonmonotonic, one resistance value corresponds to two concentrations. Both ambiguities can be avoided only if the null point and its dependence on temperature and device history are known.

Here, we use atomic layer deposition (ALD) to grow Pd nanoparticle networks on macroporous silicon. Pd initially nucleates as islands that coalesce as the number of ALD cycles increases~\cite{elam2006aldpd,weber2014aldpd,Lausecker_Munoz-Rojas_Weber_2024,zou_atomic_2021}. The cycle count therefore controls the network's proximity to the percolation threshold. Macroporous silicon provides a rigid three-dimensional scaffold with micrometer-scale pores, open gas access, and a large internal surface area~\cite{lehmann1990macroporous,scheen_palladium_2014,brinker2024mottschottky}. Pd ALD coats both the external surface and the pore walls (Figures~\ref{fig:sem} and SF\,1). The availability of this scaffold as a free-standing membrane with through-pores also offers a route to flow-through sensors with Pd-coated pore walls. We measure the response of a network annealed at 200~\degC{} to 0.05--5~\volpc{} \ce{H2} in \ce{N2} at operating temperatures of 25, 40, and 55~\degC{}. A thermodynamically informed two-channel model relates hydrogen uptake to electron scattering within the particles and improved conduction across their junctions. Fitted jointly to 300 exposure cycles, the model predicts how the null point shifts with operating temperature. To assess the effect of thermal history, we compare calibration curves before and after this anneal at each operating temperature and examine the room-temperature polarity pattern in an independently treated device. The measured curves identify the concentration ranges over which a resistance change gives an unambiguous reading. Because the null point shifts predictably with temperature, response polarity becomes an operating parameter. The operating temperature could then be chosen to place the null point at a target hydrogen concentration, such as an alarm setpoint.
\section{Results and Discussion}

\subsection{ALD-grown nanoparticle networks on macroporous silicon}

\begin{figure*}[tp]
\centering
\includegraphics[width=\linewidth]{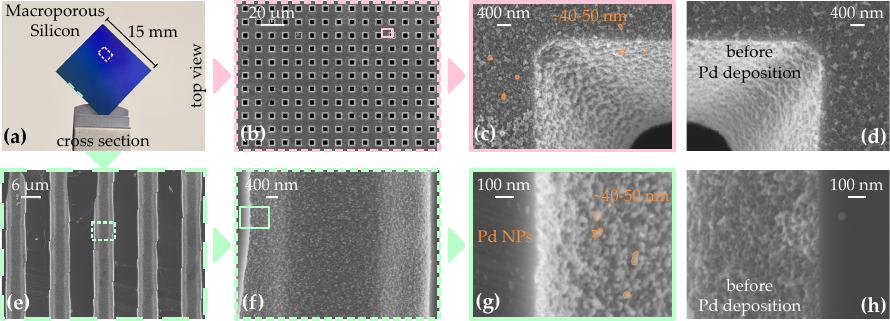}
\caption{Pd nanoparticle network grown on macroporous silicon by 1100 ALD cycles. (a) Photograph of the 15~mm specimen. (b,c) Top-view SEM images at increasing magnification and (d) an image of the uncoated top surface. (e--g) Cross-sectional SEM images of the coated pore walls and (h) an image of an uncoated pore wall. Pink and green frames indicate successive magnifications. Orange outlines identify representative Pd features with diameters of 40 to 50~nm.}
\label{fig:sem}
\end{figure*}

Macroporous silicon with 6~$\mu$m pores arranged on a 12~$\mu$m square lattice and a pore depth of 215~$\mu$m served as the scaffold. Palladium was deposited by thermal ALD from Pd(hfac)$_2$ and ozone at 200~\degC{} (Methods). Scanning electron microscopy of an 1100-cycle coating shows granular Pd on the planar top surface, around the pore openings, and on the pore walls, while the macropores remain open (Figure~\ref{fig:sem}). Individual Pd features measure 40 to 50~nm in diameter, and an analysis of 83 features yields a mean size of $42.4\pm10.3$~nm (Figure~SF\,2a,c). Energy-dispersive X-ray maps of an as-deposited 580-cycle device confirm the presence of Pd in both top view and cross section (Figure~SF\,1). The sensing devices studied below received 580 cycles. The treated device, used for the temperature study, was imaged after thermal treatment. The 118 Pd features analyzed have a mean projected-area-equivalent diameter of $29.8\pm5.7$~nm and a median of 29.0~nm (Figure~SF\,2b,d). This distribution provides the morphological input for the model.

\subsection{As-deposited networks respond positively}

\begin{figure*}[tp]
\centering
\includegraphics[width=.99\linewidth]{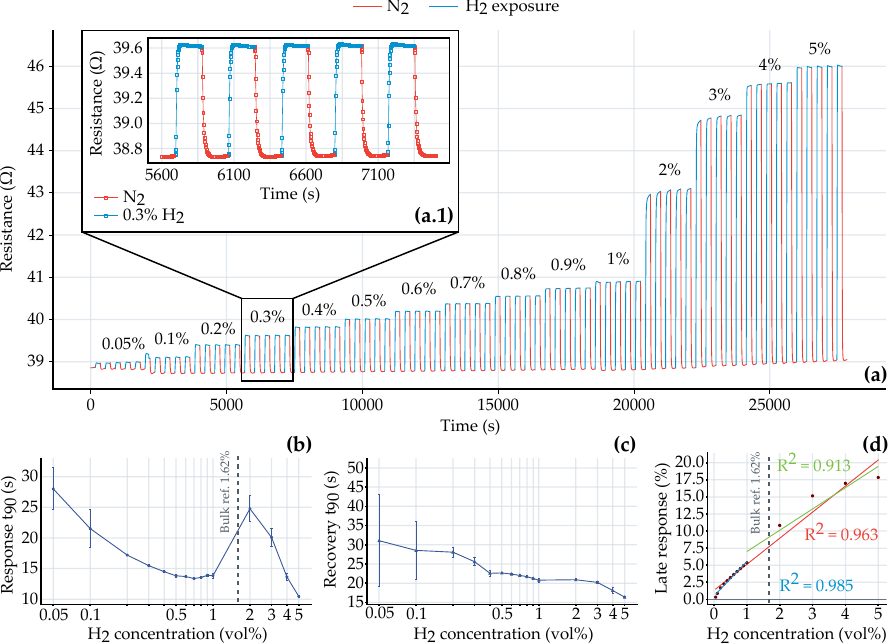}
\caption{Positive hydrogen response of the reference device (580 ALD cycles, as-deposited) at room temperature. (a) Resistance during five 180~s \ce{H2} exposures and 180~s \ce{N2} recovery periods at each concentration between 0.05 and 5~\volpc. Blue and red indicate the gas phases. The transients in (a) were smoothed, which removes the short spikes visible in the raw data (Figure~SF\,3). We used the unsmoothed data to determine $R_{\mathrm{late}}$, all response values, and $t_{90}$. The inset shows the cycles at 0.3~\volpc{}. (b,c) Response and recovery times $t_{90}$, defined as the first crossing of 90~\% of the signed change toward the late level (Methods). (d) Late response $100(R-R_0)/R_0$, where $R_0$ is the local pre-exposure baseline. Error bars indicate cycle standard deviations ($n=5$). Blue, green, and red lines show linear fits over 0.05 to 1, 1 to 5, and 0.05 to 5~\volpc{}, respectively. Dashed markers in (b,d) indicate the bulk Pd--H plateau concentration of 1.62~\volpc{} at 25~\degC.}
\label{fig:response}
\end{figure*}

Figure~\ref{fig:response}a shows the resistance of the reference device, an as-deposited 580-cycle network, during five 180~s exposures at each concentration between 0.05 and 5~\volpc{} \ce{H2} in \ce{N2}, each followed by 180~s of \ce{N2}. The resistance increases during every exposure and recovers in \ce{N2}, whereas the uncoated support exhibits no measurable response. The late response, defined as the relative resistance change averaged over 140 to 160~s after the gas switch (Methods), increases from 0.33~\% at 0.05~\volpc{} to 17.85~\% at 5~\volpc{} (Figure~\ref{fig:response}d). The calibration curve is approximately linear below 1~\volpc{} and flattens above 3~\volpc. The three linear fits compare the low- and high-concentration regimes with a fit over the full range; no single straight line captures both regimes and their crossover near the bulk plateau concentration. This behavior is consistent with competing contributions to the response, as represented by the two-channel model below. The response and recovery times $t_{90}$ are shortest at 5~\volpc{}, at 10 and 16~s, respectively (Figure~\ref{fig:response}b,c). Both exhibit a local maximum near 2~\volpc. The concentration of this maximum coincides with the bulk Pd--H plateau concentration at 25~\degC, which is 1.62~\volpc{} at 1~bar~\cite{Wicke1991,flanagan1991palladium} (Supporting Information, Section~SS\,2). Slower kinetics during $\alpha$--$\beta$ conversion are well documented for Pd nanowires~\cite{yang2010joule} and nanoparticles~\cite{langhammer2010sizekinetics}. The baseline also drifts upward during cycling at concentrations above 1 to 2~\volpc{}, and the offset persists when low concentrations are revisited (Figure~SF\,3). The persistence of this offset over the experimental period is consistent with changes in the network during repeated hydride formation and decomposition. A second as-deposited device from a separate ALD run reproduces the positive calibration curve, with responses of 0.16~\% at 0.05~\volpc{} and 16.18~\% at 5~\volpc{} (Figure~SF\,4). Its higher baseline resistance, 200 to 350~$\Omega$ compared with 39 to 46~$\Omega$, may reflect a thinner coating deposited in that run.

\subsection{Temperature shifts the null point and makes the calibration nonmonotonic}

\begin{figure*}[tp]
\centering
\includegraphics[width=.99\linewidth]{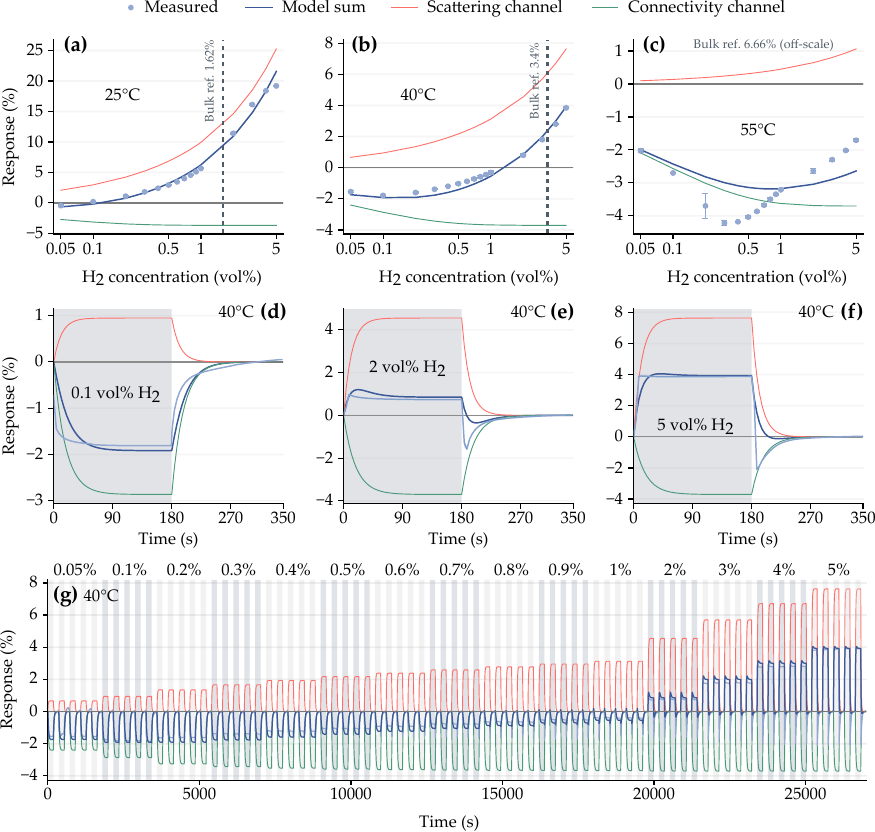}
\caption{Temperature-dependent response of the treated device (580 ALD cycles, after annealing) at three temperatures and the joint two-channel model. (a--c) Late responses at nominal temperatures of 25, 40, and 55~\degC{} (points with cycle standard deviations; $n=5$, 10, and 5, respectively). Blue, red, and green curves show the model sum, scattering channel, and connectivity channel, respectively. Dashed markers indicate the bulk Pd--H plateau concentrations of 1.62, 3.40, and 6.66~\volpc{}; the last lies outside the plotted range. (d--f) Mean transients at 40~\degC{} for 0.1, 2, and 5~\volpc{} \ce{H2}, with standard-deviation bands ($n=5$) and model curves. (g) First run at 40~\degC{}, comprising five 180~s exposure and 180~s recovery cycles per concentration, shown on a concatenated timeline. Shaded bands indicate \ce{H2} exposure. The model was fitted jointly to all 300 cycles at the three temperatures using nine parameters.}
\label{fig:model}
\end{figure*}

Following the 10~min anneal in air at 200~\degC{}, we measured the network at nominal operating temperatures of 25, 40, and 55~\degC{}, with 75, 150, and 75 hydrogen exposure cycles, respectively. Figure~\ref{fig:model}a--c shows the late responses. At 25~\degC{}, the response is positive above approximately 0.1~\volpc{} and slightly negative below this concentration. At 40~\degC{}, the response is negative up to approximately 1~\volpc{} and positive at higher concentrations, reaching 3.8~\% at 5~\volpc{}. At 55~\degC{}, the response is negative throughout the measured range, reaching $-4.2$~\% at 0.3~\volpc{} before becoming less negative as concentration increases. The calibration curve at 55~\degC{} is therefore nonmonotonic. At this temperature, the same late-response value can correspond to two hydrogen concentrations, one on either side of the minimum.

The transient responses reflect the same changes. At 40~\degC{}, exposure to 0.1~\volpc{} produces a monotonic decrease in resistance (Figure~\ref{fig:model}d). Exposure to 2~\volpc{} produces a rapid decrease followed by a slower increase, and the resistance undershoots the baseline during recovery (Figure~\ref{fig:model}e). At 5~\volpc{}, the positive branch dominates, but the recovery undershoot persists (Figure~\ref{fig:model}f). These traces are consistent with two contributions of opposite sign and different kinetics. The thick-coating device (1100 ALD cycles, as-deposited) exhibits the same trend between room temperature and 30~\degC{}, with pronounced overshoots at the lowest concentrations and a strongly reduced kinetic maximum (Figure~SF\,6).

\subsection{Two electrical channels driven by one hydrogen loading}

We describe all three temperature programs using a single model that separates hydrogen uptake from electrical transduction. The model describes the response of the treated device as the sum of two opposing electrical contributions from hydrogen absorption,
\begin{equation}
S_{\mathrm{mod}}(t,T)=S_{\mathrm{s}}(t,T)+S_{\mathrm{p}}(t,T).
\label{eq:total-main}
\end{equation}
$S_{\mathrm{s}}$ is the positive scattering contribution (subscript s), and $S_{\mathrm{p}}$ is the negative connectivity contribution (subscript p, for percolation). The latter represents improved conductive paths as expansion reduces the separation of critical junctions. Both contributions are fractional changes relative to the zero-loading resistance $R_{\mathrm{ref}}(T)$, such that $R_{\mathrm{mod}}/R_{\mathrm{ref}}=1+S_{\mathrm{mod}}$. Their equilibrium targets depend on the same inferred hydrogen loading, but the channels evolve separately toward those targets. Their sum is compared with the measured response after cycle-local normalization. The individual channel curves represent components assigned by the model.

The equilibrium loading $x$ of a particle with diameter $D$ follows a Lacher-type isotherm~\cite{lacher1937solubility},
\begin{equation}
\tfrac{1}{2}\ln\!\left(\frac{p_{\ce{H2}}}{p_0}\right)=\ln\!\left(\frac{x_D}{r-x_D}\right)-\frac{E_{\mathrm{eff}}x_D}{RT}+\frac{E_0(T)+u_{\mathrm{surf}}/D}{RT},
\label{eq:lacher-main}
\end{equation}
where $r=0.6$ is the maximum site occupancy, $E_{\mathrm{eff}}$ is an effective H--H interaction, $E_0(T)$ is an insertion energy incorporating the Pd--H entropy of $-94.5$~J\,K$^{-1}$\,mol$^{-1}$~\ce{H2}, and $u_{\mathrm{surf}}/D$ is a surface term accounting for the size dependence of nanoparticle isotherms~\cite{griessen2016thermodynamics}. The loading is averaged over the measured size distribution shown in Figure~SF\,2d. This loading drives two electrical channels. The scattering channel follows the tabulated resistivity of bulk PdH$_x$~\cite{otterson1969absorption},
\begin{equation}
S_{\mathrm{s}}^{\mathrm{eq}}=A_{\mathrm{s}}(T)\left[\frac{\rho(x)}{\rho(0)}-1\right].
\label{eq:scatter-main}
\end{equation}
Here, $\rho(x)/\rho(0)$ is the relative resistivity of bulk PdH$_x$ interpolated from tabulated data, $A_{\mathrm{s}}(T)$ scales this intrinsic response to that of the network, and $S_{\mathrm{s}}^{\mathrm{eq}}$ is the equilibrium target of the scattering channel.
The connectivity channel represents a finite population of critical junctions that close as the particles expand,
\begin{equation}
S_{\mathrm{p}}^{\mathrm{eq}}=-P_{\max}\left[1-\exp\!\left(-\frac{\eta_{\mathrm{H}}\overline{D}}{d^{\ast}}\,x\right)\right],
\label{eq:perc-main}
\end{equation}
where $\eta_{\mathrm{H}}=0.063$ per H/Pd is the linear expansion coefficient, $\overline{D}$ is the mean diameter, and $d^{\ast}$ is an effective gap scale. The two channels relax toward their respective targets with first-order time constants $\tau_{\mathrm{s}}(T)$ and $\tau_{\mathrm{p}}$. These effective time constants summarize contributions from surface exchange, transport, and network rearrangement, processes that differ even between facets and vertices of individual Pd nanocrystals~\cite{johnson_facets_2019}. The measured signal is the sum of the two channels after applying the same cycle-local baseline normalization used for the data. Temperature enters through $E_0(T)$, $A_{\mathrm{s}}(T)$, and $\tau_{\mathrm{s}}(T)$. All other parameters are shared across the three temperatures. Nine parameters were fitted to the 300 cycles using robust nonlinear least squares. The full model specification, fitting procedure, cycle-bootstrap confidence intervals, and holdout tests are provided in Sections~SS\,5 to SS\,7 of the Supporting Information.

Across all 300 exposure cycles at 25, 40, and 55~\degC{}, the joint fit gives a pointwise root-mean-square error of 0.57 percentage points (Figure~\ref{fig:model} and Figures~SF\,8 to SF\,11). It captures the temperature-dependent response polarity and, at 40~\degC{}, the initial resistance drop, slower rise, and recovery undershoot (Figure~\ref{fig:model}d--f). Table~\ref{tab:params} lists the fitted parameters. The connectivity channel saturates at a resistance decrease of 3.7~\%, with a time constant of 17.8~s at all temperatures. The scattering amplitude decreases from $A_{\mathrm{s}}=2.56$ at 25~\degC{} to 1.06 at 40~\degC{} and 0.19 at 55~\degC{}, while the channel's time constant decreases from 25.8 to 5.9~s. The equilibrium null point of the total response shifts from 0.12~\volpc{} at 25~\degC{} to 1.36~\volpc{} at 40~\degC{} and beyond the measured range at 55~\degC{}. We conclude from the joint fit that the sign reversal results from a decrease in the positive scattering contribution rather than an increase in the negative connectivity contribution.

\begin{table}[tp]
\centering
\caption{Parameters of the joint two-channel fit to the treated device. Temperature-dependent quantities are listed separately for each temperature. The shared parameters are $P_{\max}=3.70$~\%, $\tau_{\mathrm{p}}=17.8$~s, $d^{\ast}=0.113$~{\AA}, and $E_{\mathrm{eff}}=10$~kJ\,mol$^{-1}$~H (lower bound). The null point is the modeled equilibrium concentration at which the total response changes sign. The value at 55~\degC{} is extrapolated beyond the measured range. Cycle-bootstrap intervals are provided in Table~ST\,2.}
\label{tab:params}
\begin{tabular}{lcccc}
\toprule
$T$ (\degC) & $E_0(T)$ (kJ\,mol$^{-1}$~H) & $A_{\mathrm{s}}(T)$ & $\tau_{\mathrm{s}}(T)$ (s) & Null point (\volpc{} \ce{H2}) \\
\midrule
25 & 0.94 & 2.56 & 25.8 & 0.12 \\
40 & 1.65 & 1.06 & 11.9 & 1.36 \\
55 & 2.36 & 0.19 & 5.9  & 26 \\
\bottomrule
\end{tabular}
\end{table}

At 55~\degC{}, the model captures the nonmonotonic response, with a fitted minimum of $-3.2$~\% near 0.8~\volpc{}. The measured minimum occurs at 0.3~\volpc{} and is about one percentage point deeper ($-4.2$~\%). The shared connectivity amplitude $P_{\max}$ limits the depth of the fitted minimum. The concurrent 6~\% baseline increase (Figure~SF\,7) suggests that the contribution of individual junctions to the network resistance may change during cycling. At 25~\degC{}, the measured calibration curve flattens above the bulk plateau pressure, and the model overestimates the response at 5~\volpc{} by 1.75 percentage points. Partial conversion to the $\beta$ phase could contribute to this flattening through saturation of the resistivity increase~\cite{otterson1969absorption}.

\subsection{The scattering amplitude decreases faster than the hydrogen loading}

The most pronounced parameter trend is the temperature dependence of the scattering amplitude. $A_{\mathrm{s}}$ decreases by a factor of 13.3 between 25 and 55~\degC{}. Expressed as an Arrhenius slope, this decrease corresponds to an apparent energy of approximately 70~kJ\,mol$^{-1}$. The Pd--H thermodynamics in Equation~\ref{eq:lacher-main} already reduce the equilibrium loading at fixed \ce{H2} pressure by a factor of 1.6 over the same temperature interval. The resistivity of bulk PdH$_x$ at fixed $x$ is nearly independent of temperature~\cite{flanagan1991palladium}. The scattering channel of this network therefore loses amplitude much faster than can be explained by the dilute lattice solution of hydrogen in Pd. This temperature dependence is characteristic of hydrogen in strongly binding sites. Chemisorbed hydrogen on Pd surfaces has a binding energy of approximately 45~kJ\,mol$^{-1}$~H~\cite{conrad1974adsorption}, while hydrogen at grain boundaries and dislocations in nanocrystalline Pd has a trapping energy of 20 to 60~kJ\,mol$^{-1}$~H relative to lattice sites~\cite{kirchheim1988trapping,pundt2006hydrogen}. In a network of 30~nm particles, a large fraction of Pd atoms is located at surfaces and internal boundaries. Hydrogen at these sites occupies the interfaces where conduction electrons are scattered as the particle size approaches the electron mean free path. We therefore attribute the scattering channel to hydrogen at boundaries, whose population decreases steeply with increasing temperature, rather than to the lattice solution alone. The time constant of this channel decreases with an apparent activation energy of 40~kJ\,mol$^{-1}$ (Table~ST\,2), within the range reported for hydrogen absorption and desorption in Pd nanoparticles~\cite{stolas2020surfactants,alekseeva2021grain,langhammer2010sizekinetics}.

\subsection{Sub-{\aa}ngstr{\"o}m gap closure and suppression of the miscibility gap}

The connectivity channel converts hydrogen loading into a decrease in resistance on a length scale of $d^{\ast}=0.11$~{\AA}. A particle of 30~nm expands by 0.19~{\AA} at $x=0.01$ and by 1.9~{\AA} at $x=0.1$. Within the model, this channel is therefore half-saturated at a loading of only approximately 0.004 and fully saturated well within the $\alpha$ phase. This model behavior explains why the negative response dominates at low concentrations and why the sign reversal occurs at the low-concentration end of the calibration curve. The value of $d^{\ast}$ is smaller than the decay length of tunneling conductance across a vacuum gap, approximately 0.5~{\AA} for a Pd work function of 5~eV~\cite{simmons1963tunneling}. This effective scale favors an interpretation in terms of threshold switching rather than smooth changes in tunneling conductance. In this interpretation, the first increments of expansion close a threshold population of nearly touching junctions, as in nanogap and break-junction hydrogen switches~\cite{favier2001mesowire,lee2011nanogap,kiefer2008nanotrench}. Within the model, saturation at $P_{\max}=3.7$~\% quantifies the fraction of the treated device's network resistance assigned to these junctions. The invariance of this saturation amplitude between 25 and 55~\degC{} reflects the model's treatment of the junction population as a structural property, with temperature affecting the connectivity channel only through hydrogen loading.

The isotherm parameter provides further insight into nanoscale effects. In the Lacher description, the bulk critical temperature of approximately 270~\degC{} corresponds to $E_{\mathrm{eff}}\approx30$~kJ\,mol$^{-1}$~H. The fit places $E_{\mathrm{eff}}$ at its lower bound of 10~kJ\,mol$^{-1}$~H, where the miscibility gap closes far below room temperature. The network thus behaves as though $\alpha$--$\beta$ coexistence were strongly suppressed. Reduced particle size, surface stress, and elastic clamping by a rigid substrate all narrow the miscibility gap of Pd nanoparticles~\cite{yamauchi2008nanosize,griessen2016thermodynamics,syrenova2015hysteresis,baldi2014insitu,dyck_phase_2024,wagner2010resistivity}. Particles grown by ALD on a silicon scaffold are clamped on one side, converting part of the hydrogen-induced strain into stress and raising the chemical potential of hydrogen. The absence of plateaus in the calibration curves at 40 and 55~\degC{} is consistent with this interpretation. The kinetic maximum near 2~\volpc{} at room temperature and the residual at 25~\degC{} and 5~\volpc{} are consistent with possible phase conversion in part of the particle population at the highest concentrations. A single-branch isotherm remains the simplest description that captures the three temperature programs, but its effective parameters do not explicitly resolve phase conversion.

\subsection{Implications for calibration and operation}

The two-channel description has direct implications for the use of nanogranular Pd sensors. First, the sign of the response depends on the operating point. A treated device of this type exhibits a negative response below and a positive response above a null point that shifts from 0.1~\volpc{} at 25~\degC{} to above 5~\volpc{} at 55~\degC{}. Second, the calibration curve at 55~\degC{} is nonmonotonic. Each late response above the minimum at 0.3~\volpc{} occurs at two concentrations, one on either side of the minimum. A single late-response value therefore cannot identify the concentration unless the monotonic branch is known. Third, the transient response resolves this ambiguity. The two channels have different time constants, 6 to 26~s for scattering and 18~s for connectivity, so the initial slope and recovery undershoot reflect the balance between the channels (Figure~\ref{fig:model}d--f). Calibration using the full transient response is therefore more robust against sign ambiguity than calibration using a single late value. The physical principles apply to Pd networks near percolation in general; the porous scaffold adds a rigid three-dimensional support compatible with flow-through geometries.

Table~\ref{tab:benchmark} compares this work with Pd architectures of known response polarity. Continuous films exhibit positive responses. Break junctions, nanogaps, and discontinuous films exhibit negative responses and are often fast, although they exhibit hysteresis. Nanoparticle arrays and near-percolation films exhibit both signs, depending on morphology. In the ALD network on macroporous silicon, temperature shifts the balance between scattering and connectivity, and the two-channel model describes the resulting polarity changes across the measured concentration range. Measurements extend down to 0.05~\volpc{}, and the response time is 10~s at 5~\volpc{}, making the performance competitive with that of resistive Pd sensors without catalytic coatings~\cite{darmadi2020review}. Faster and more sensitive Pd sensors are available, particularly plasmonic and single-nanowire devices~\cite{nugroho2019hybrid,yang2010smaller,offermans2009nanowire,Jo_Kim_Lee_Kim_Yoo_Choi_Kim_Kwon_Yoo_Yang_et_al._2023}. The contribution of the present architecture lies in its scalable three-dimensional scaffold and an interpretable response that allows calibration under changing temperatures.

\begin{table}[tp]
\centering
\caption{Polarity of the resistive hydrogen response of Pd architectures near room temperature. A positive response denotes an increase in resistance during \ce{H2} exposure.}
\label{tab:benchmark}
\small
\begin{tabular}{>{\raggedright\arraybackslash}p{5.2cm}>{\raggedright\arraybackslash}p{3.2cm}>{\raggedright\arraybackslash}p{3.4cm}l}
\toprule
Architecture & Dominant mechanism & Response sign & Refs. \\
\midrule
Continuous films and thick wires & lattice resistivity & positive & \cite{hubert2011review,yang2010smaller} \\
Mesowire arrays with break junctions & gap closure & negative & \cite{favier2001mesowire} \\
Ultrathin discontinuous films & gap closure & negative & \cite{xu2005ultrathin,kiefer_transition_2010} \\
Nanogaps on elastomers & gap closure & negative & \cite{lee2011nanogap} \\
Nanoparticle arrays and near-percolation films & both & morphology dependent & \cite{ramanathan_crossover_2010,xie_response_2012,Das2020} \\
ALD network on macroporous Si (this work) & both & switched by $T$ and device history & -- \\
\bottomrule
\end{tabular}
\end{table}

\subsection{Effect of annealing on the hydrogen response}

\begin{figure}[H]
\centering
\includegraphics[width=.99\linewidth]{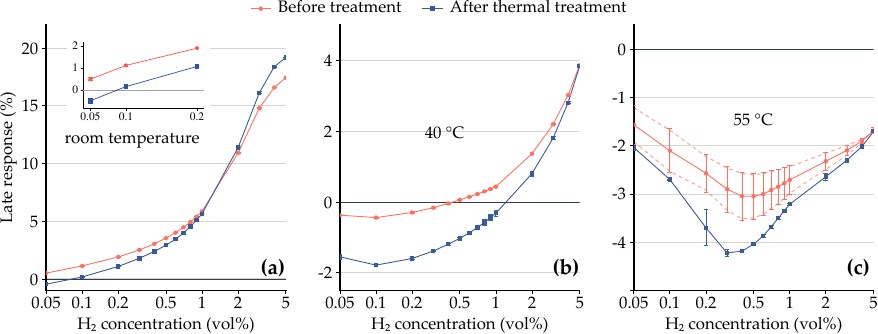}
\caption{Effect of annealing on the hydrogen response of the same 580-cycle Pd network. Late response after 90 days of storage (red circles) and after subsequent annealing in air for 10~min at 200~\degC{} (blue squares), measured at (a) room temperature, (b) 40~\degC{}, and (c) 55~\degC{}. The inset in (a) expands the low-concentration region around zero response. Each point represents the mean of $100(R-R_0)/R_0$ over 140 to 160~s after \ce{H2} onset. Error bars indicate cycle standard deviations, with $n=10$ for the aged state at 55~\degC{} and the treated state at 40~\degC{}, and $n=5$ otherwise. Pale-red dashed curves distinguish the two runs for the aged state at 55~\degC{}.}
\label{fig:treatment}
\end{figure}

To determine how thermal history affects the calibration at a fixed operating temperature, we compared the same network at nominal temperatures of 25, 40, and 55~\degC{} before and after annealing (Figure~\ref{fig:treatment}). The first measurements followed 90 days of storage after deposition. The subsequent anneal in air at 200~\degC{} for 10~min followed a protocol adapted from Pd nanowire sensors~\cite{Kim2024}.

At room temperature, the response was positive throughout the measured range before annealing. At 0.05~\volpc{} \ce{H2}, it changed from $+0.51\pm0.06$~\% before treatment to $-0.46\pm0.12$~\% afterward (Figure~\ref{fig:treatment}a, inset). At 5~\volpc{}, it remained positive, increasing from 17.44 to 19.19~\%. A second device subjected to the same anneal reproduced this polarity pattern in two successive runs. Its responses were $-0.59\pm0.08$ and $-0.19\pm0.03$~\% at 0.05~\volpc{} and 18.56 and 17.78~\% at 5~\volpc{}, respectively (Figure~SF\,5).

At 40~\degC{}, both calibration curves crossed zero, but the crossing shifted to a higher concentration after annealing (Figure~\ref{fig:treatment}b). The measured crossing lay between 0.4 and 0.5~\volpc{} before treatment and between 1 and 2~\volpc{} afterward. At 0.1~\volpc{}, the response changed from $-0.44\pm0.01$ to $-1.78\pm0.04$~\%. At 55~\degC{}, both states responded negatively throughout the measured range. Annealing deepened the minimum, and at 0.5~\volpc{} the response changed from $-3.05$ to $-4.04$~\% (Figure~\ref{fig:treatment}c).

Several processes could shift the balance between scattering and junction conduction after an air anneal. Particle coarsening or relaxation of growth stress could change the gaps along conducting paths and hence the resistance change produced by particle expansion~\cite{kiefer_transition_2010}. Surface oxidation during the anneal and reduction during subsequent hydrogen exposure could also change the surface state~\cite{zemlyanov2006pdoxidation}, while annealing of cycling-induced defects could alter the hydriding kinetics~\cite{delmelle_effect_2015,alekseeva2021grain}. The paired calibrations establish that the effect of treatment persists at the measurement temperatures. Distinguishing these mechanisms would require structural or surface measurements before and after treatment.
\section{Conclusions and Perspective}

Changing the operating temperature can reverse the sign of the resistance response to hydrogen in a nanogranular Pd network on macroporous silicon, even at a fixed hydrogen concentration. In the two-channel model, this reversal reflects the balance between two effects of hydrogen uptake. Electron scattering raises the resistance within the particles, while particle expansion narrows the gaps between them and lowers the resistance at their junctions. Between 25 and 55~\degC{}, the fitted scattering amplitude decreases thirteenfold, allowing sub-{\aa}ngstr{\"o}m changes in gap width to dominate the response.

For the annealed network, the model predicts that the equilibrium null point shifts from 0.12~\volpc{} \ce{H2} at 25~\degC{} to 1.4~\volpc{} at 40~\degC{}, crossing the reported alarm setpoints of 0.4 and 1.0~\volpc{}. The 200~\degC{} anneal also altered the calibration at each operating temperature. It reversed the low-concentration response at 25~\degC{}, shifted the null point to a higher concentration at 40~\degC{}, and deepened the negative response minimum at 55~\degC{}.

At the equilibrium null point, the two contributions cancel and the sensor gives the same resistance reading as in hydrogen-free gas. When the contributions evolve on different time scales, the transient response can still indicate hydrogen uptake. At 55~\degC{}, the measured response remained negative throughout the tested range but passed through a minimum near 0.3~\volpc{}, so two concentrations on opposite sides of it could produce the same late resistance change. Quantitative sensing from the late response therefore requires a defined thermal pretreatment and an operating temperature at which the intended concentration range lies on a monotonic branch of the calibration curve. In such a network, response polarity is an operating parameter rather than a fixed property. Tuning the operating temperature could place the null point at a chosen hydrogen concentration, such as an alarm setpoint, so that exposures below and above that concentration produce resistance changes of opposite sign.

The model infers hydrogen loading from the electrical response. Measuring equilibrium uptake and resistance together on the annealed network under \ce{H2} and \ce{D2} at 40~\degC{} would provide an independent test. If isotope substitution shifts the null point mainly through its effect on uptake, cancellation should occur at different partial pressures but at similar atomic loadings (H/Pd and D/Pd). Because the calibration was measured with \ce{N2} as the carrier gas, its applicability to leak detection should next be tested in humid air. For reactor monitoring, Pd-coated silicon membranes with through-pores offer a way to test how liquid-phase transport and competing adsorbates affect the calibration under flow.

The same competition may arise in other nanoparticle networks where absorption increases particle resistivity while expansion improves conduction across interparticle junctions. In such materials, temperature could likewise select which effect dominates and thereby set the sign of the response. Particle deformation, moreover, does not require absorption into the bulk. Physisorption of gases and vapors, including water, induces reversible strains in nanoporous solids that are captured by surface-stress and solvation-pressure models. These strains can even change sign along the isotherm, with contraction during capillary condensation and expansion at higher filling~\cite{Gor2017, Detsi2011}. Adsorbates also modulate the resistance of nanoporous gold by surface scattering, for example upon water adsorption~\cite{Wahl2010, Hakamada2016}. In ligand-capped gold nanoparticle films, by contrast, vapor-induced swelling of the matrix between the particles widens the tunneling gaps and raises the resistance~\cite{Olichwer2016}. How adsorption-induced strain of the particles themselves, acting on the interparticle contacts, competes with surface scattering inside them has, to our knowledge, not yet been resolved. This question is the adsorption counterpart of the competition reported here. Resolving this interplay would link the mechanics of sorption at nanoscale interfaces to charge transport in nanostructured matter. It would also turn the coupling between sorption, strain, and conduction from a source of ambiguity into a design principle for sensors, switches, and adaptive nanomaterials well beyond palladium--hydrogen.
\section{Methods}

\subsection{Substrates and atomic layer deposition}
Macroporous silicon (SmartMembranes GmbH) consisted of n-doped single-crystal Si(100) with a square array of pores measuring 6~$\mu$m in diameter and 215~$\mu$m in depth, with a center-to-center spacing of 12~$\mu$m. Pieces measuring $15\times15$~mm$^2$ were used as received. Palladium was deposited by thermal ALD in a home-built reactor using palladium(II) hexafluoroacetylacetonate (Pd(hfac)$_2$, minimum purity 95~\%, Strem Chemicals) and ozone generated from pure oxygen~\cite{zou_atomic_2021,cheng_adsorption_2021}. The substrate and precursor temperatures were 200 and 65~\degC, respectively. Each ALD cycle consisted of two 30~s Pd(hfac)$_2$ pulses, each followed by a 90~s purge, and one 1~s ozone pulse followed by a 60~s purge. Five devices were examined. Four devices received 580 ALD cycles: the as-deposited reference device, the second as-deposited device from a separate run, the treated device used for the aging and temperature measurements, and the replication device. The as-deposited thick-coating device received 1100 ALD cycles. The treated device was first measured approximately 90 days after deposition and was subsequently heated on a hot plate at 200~\degC{} for 10~min in ambient air~\cite{Kim2024}. The replication device received the same treatment.

\subsection{Electron microscopy}
SEM images and EDS maps were acquired at 15~kV. Feature sizes were determined by manually outlining Pd features. A total of 83 features were evaluated for the 1100-cycle coating. For the treated device, 118 features were outlined in an image in which the 200~nm scale bar spanned 151 pixels. Each feature was assigned the diameter of a circle with the same projected area. Further details, including the EDS acquisition, are provided in Section~SS\,1.

\subsection{Hydrogen exposure and resistance measurement}
Two mass-flow controllers (MC series, Alicat Scientific) mixed a 5~\volpc{} \ce{H2} mixture in \ce{N2} with pure \ce{N2} to obtain \ce{H2} concentrations ranging from 0.05 to 5~\volpc{}. The gas passed through a 7.05~mL chamber at a total flow rate of 1 standard liter per minute and a pressure near 1~bar (Figure~SF\,12). The ideal exchange time for a well-mixed chamber, $t_{90}=\ln(10)\,V/Q$, is 1~s. Four tungsten-carbide pins contacted the sample. Resistance was measured in a four-point configuration using a Keithley 6220 current source supplying alternating $\pm1$~mA and a Keithley 2182 nanovoltmeter. The median sampling intervals were 0.61~s for the initial calibration and 1.11~s for the post-treatment data. A PT100 located at a sample corner measured the temperature and provided feedback for a Kapton heater. A thermal-conductivity hydrogen sensor (Infineon TCIXTMA1) and a pressure sensor (Bosch BMP581) recorded the reference gas and pressure signals, respectively, for phase alignment. Each measurement cycle consisted of 180~s of \ce{H2} exposure followed by 180~s of recovery in \ce{N2}. The measurement program began with five conditioning cycles at 5~\volpc{}, followed by an ascending concentration series with five cycles per concentration. The uncoated support showed no measurable response. The median PT100 temperatures for the programs designated room temperature (or 25~\degC), 40~\degC, and 55~\degC{} were 23.7, 40.0, and 55.0~\degC, respectively. The thick-coating device was compared at 24.0 and 30.0~\degC.

\subsection{Response definitions}
Responses were referenced to the local \ce{N2} baseline, $R_0$, defined as the median resistance during the interval from 40 to 10~s before \ce{H2} onset. The late response is $100(R_{\mathrm{late}}-R_0)/R_0$, where $R_{\mathrm{late}}$ is the mean resistance from 140 to 160~s after onset. The response time, $t_{90}$, is the first time at which the signed resistance change from $R_0$ reaches 90~\% of the change from $R_0$ to $R_{\mathrm{late}}$. The recovery time is defined analogously, using the resistance at switch-off as the starting value and the mean resistance from 140 to 160~s after the switch to \ce{N2} as the endpoint. Gas-switch times were identified from the first sustained change in the reference sensor signal. For recordings without telemetry and for the aged room-temperature and 40~\degC{} recordings with lagging reference-sensor transitions, gas-switch times were instead identified from the largest absolute derivative of the smoothed resistance trace near the programmed switch (Section~SS\,8). Experimental uncertainties are cycle standard deviations calculated from five cycles per concentration in each run, or ten cycles when two runs are pooled, as specified in the captions. Uncertainties in fitted parameters are reported separately as conditional 95~\% cycle-bootstrap intervals.

\subsection{Global fit}
The two-channel model was fitted jointly to all 300 post-treatment cycles of the treated device over the full 0.05 to 5~\volpc{} \ce{H2} range at nominal temperatures of 25, 40, and 55~\degC. Hydrogen loading follows Equation~\ref{eq:lacher-main}, using $r=0.6$, $\eta_{\mathrm{H}}=0.063$, $u_{\mathrm{surf}}=10$~kJ\,nm\,mol$^{-1}$~H, and the measured size distribution of the treated device represented on a grid of 15 diameters. The two channels follow Equations~\ref{eq:scatter-main} and \ref{eq:perc-main}, with first-order kinetics. The temperature dependences of $A_{\mathrm{s}}(T)$ and $\tau_{\mathrm{s}}(T)$ were parameterized using bounded Arrhenius-type relations with a reference temperature of 40~\degC. The same cycle-local baseline normalization was applied to measured and simulated cycles. The simulated states were propagated continuously throughout the measurement program, including the conditioning cycles. The objective function used a soft-$L_1$ loss on group means, with repeatability scaling and balanced weights for temperature, concentration, and gas phase, and was minimized by constrained nonlinear least squares. Confidence intervals were obtained from 200 cycle-bootstrap refits. The complete model specification, parameter bounds, residual diagnostics, and holdout tests are provided in Sections~SS\,5 to SS\,7.

\section*{Acknowledgments}
This work was funded by the Deutsche Forschungsgemeinschaft (DFG, German Research Foundation) -- SFB 1615 -- 503850735 (SMART Reactors for Future Process Engineering), project A03. P. Huber also acknowledges funding from the DFG under Germany's Excellence Strategy -- EXC 3120/1 -- 533771286 (BlueMat, Water-Driven Materials). The authors thank the Centre for Molecular Water Science (CMWS), Hamburg, Germany, for scientific exchange and support.

\section*{Conflicts of Interest}
The materials described in this research are the subject of European patent application No.~26210112.4. Niclas Alkazaz, Robert Zierold, and Patrick Huber are the inventors. Hamburg University of Technology (TUHH) will own any patent granted from this application.

\section*{Supporting information}
Electron microscopy and feature-size analysis; bulk thermodynamic reference markers; extended room-temperature recordings; device comparisons and telemetry; full model specification; complete comparisons of model predictions and measurements at each temperature; residual diagnostics; cycle-bootstrap intervals and temperature holdout tests; and measurement setup and data processing (PDF).

\section*{Data and code availability}
Measurement data and analysis code that support the findings of this study are available from the corresponding author upon reasonable request.

\providecommand{\noopsort}[1]{}\providecommand{\singleletter}[1]{#1}%
\providecommand{\latin}[1]{#1}
\makeatletter
\providecommand{\doi}
  {\begingroup\let\do\@makeother\dospecials
  \catcode`\{=1 \catcode`\}=2 \doi@aux}
\providecommand{\doi@aux}[1]{\endgroup\texttt{#1}}
\makeatother
\providecommand*\mcitethebibliography{\thebibliography}
\csname @ifundefined\endcsname{endmcitethebibliography}
  {\let\endmcitethebibliography\endthebibliography}{}

\clearpage
\setcounter{page}{1}
\renewcommand*{\thepage}{S\arabic{page}}
\setcounter{figure}{0}
\setcounter{table}{0}
\setcounter{equation}{0}
\setcounter{section}{0}
\setcounter{secnumdepth}{2}
\renewcommand*{\thefigure}{SF\,\arabic{figure}}
\renewcommand*{\thetable}{ST\,\arabic{table}}
\renewcommand*{\theequation}{SE\,\arabic{equation}}
\renewcommand*{\thesection}{SS\,\arabic{section}}
\begin{center}
{\Large\bfseries Supporting Information}\\[1ex]
{\bfseries Temperature Flips the Sign of the Hydrogen Response in Nanogranular Palladium Networks}\\[1ex]
Niclas Alkazaz, Robert Zierold, and Patrick Huber
\end{center}
\section{Electron microscopy}
\label{si:microscopy}

\begin{figure}[htbp]
\centering
\includegraphics[width=0.98\linewidth]{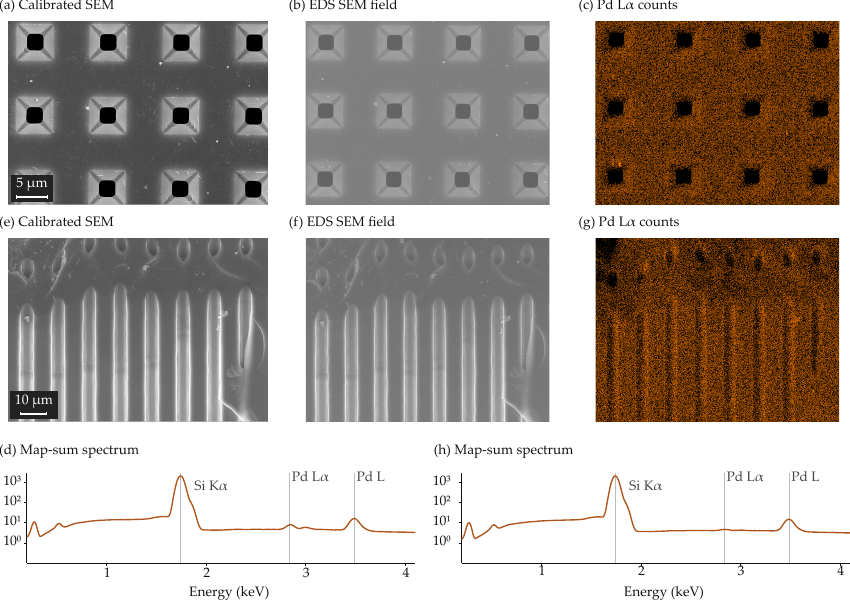}
\caption{EDS characterization of an as-deposited 580-cycle Pd ALD device in top view (a--d) and cross section (e--h). (a,e) Calibrated companion SEM images with 5 and 10~$\mu$m scale bars, respectively. (b,f) Electron images of the EDS fields. (c,g) Pd~L$\alpha$ count maps. (d,h) Map-sum spectra on logarithmic axes, expressed in counts per second per 10~eV channel, with Si and Pd line positions indicated. The companion SEM images show fields different from the EDS fields. Data were acquired at 15~kV, with live times of 1730~s and 3991~s for the top view and cross section, respectively. Map brightness represents qualitative X-ray counts.}
\label{fig:si-eds}
\end{figure}

EDS maps of the second as-deposited device were acquired at 15~kV, with spectrum live times of 1730~s for the top view and 3991~s for the cross section (Figure~\ref{fig:si-eds}). The paired electron images and elemental maps show the same fields, whereas the calibrated companion SEM images show separate fields. Map intensity represents qualitative X-ray counts rather than a corrected Pd concentration.

\begin{figure}[tp]
\centering
\includegraphics[width=\linewidth]{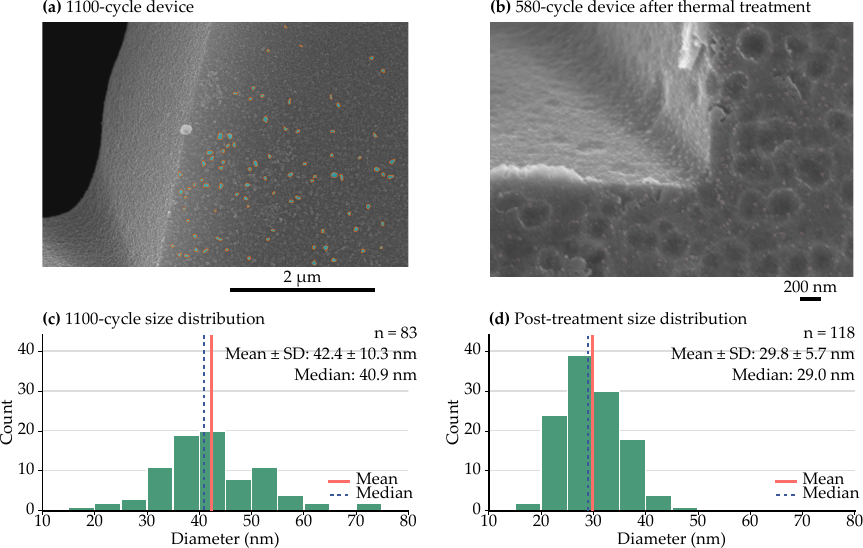}
\caption{Feature-size analysis of the Pd networks. (a,c) The 1100-cycle coating before thermal treatment: 83 features with a mean diameter of $42.4\pm10.3$~nm and a median of 40.9~nm. (b,d) The treated device after treatment in air at 200~\degC{} for 10~min: 118 manually outlined features with a mean projected-area-equivalent diameter of $29.8\pm5.7$~nm and a median of 29.0~nm. Histograms have 5~nm bins. Solid red and dashed blue lines indicate the mean and median, respectively. The distribution in (d) provides the morphological input for the two-channel model.}
\label{fig:size}
\end{figure}

The feature-size analysis in Figure~\ref{fig:size} included 83 selected features of the 1100-cycle coating and 118 manually outlined features of the treated device after thermal treatment. In the post-treatment image, the 200~nm scale bar spanned 151 pixels. For each feature, a circular diameter equivalent to its projected area was calculated from the outline midline. The midline area was estimated from the inner and outer areas to reduce sensitivity to marker width. The features of the treated device have a mean diameter of $29.8\pm5.7$~nm, a median of 29.0~nm, an interquartile range of 25.5 to 33.2~nm, and a range of 19.3 to 48.2~nm. These images characterize the sampled regions; they do not provide a quantitative coverage profile over the full pore depth.

\clearpage

\section{Bulk thermodynamic reference markers}
\label{si:bulkref}

The bulk Pd--H plateau markers in Figures~2 and 3 of the main text are calculated from
\begin{equation}
\ln\!\left(\frac{p_{\mathrm{pl}}}{p^\circ}\right)=\frac{\Delta H-T\Delta S}{RT},
\label{eq:si-plateau}
\end{equation}
using $p^\circ=1$~bar, the bulk absorption enthalpy $\Delta H=-38.4$~kJ\,mol$^{-1}$~\ce{H2}~\cite{Wicke1991,flanagan1991palladium}, and the entropy $\Delta S=-94.5$~J\,mol$^{-1}$\,K$^{-1}$~\ce{H2}, which is also used in the model. At a total pressure of 1~bar, the plateau concentrations are 1.62, 3.40, and 6.66~\volpc{} at 25, 40, and 55~\degC, respectively. The value at 25~\degC{} is a reference at the nominal temperature; the measured room-temperature median was 23.7~\degC. These markers provide bulk thermodynamic context and represent neither fitted transition pressures nor measured phase boundaries of the network. Particle size, hysteresis, and mechanical constraint shift absorption and desorption relative to bulk Pd~\cite{griessen2016thermodynamics,syrenova2015hysteresis,dyck_phase_2024}.

\section{Initial calibration, repeatability, and kinetics}
\label{si:calibration}

\begin{figure}[htbp]
\centering
\includegraphics[width=0.98\linewidth]{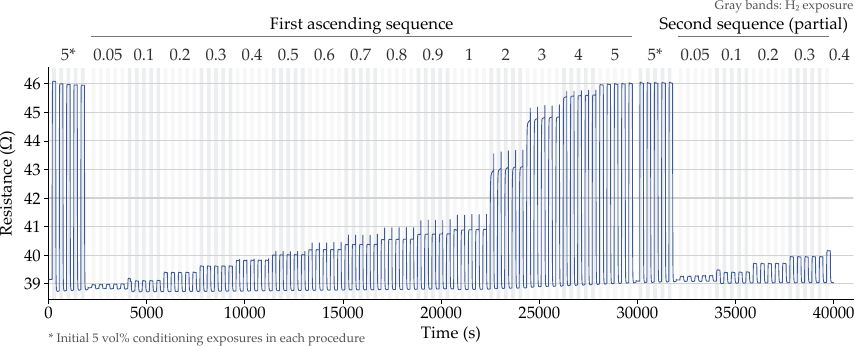}
\caption{Extended room-temperature raw resistance recording of the reference device (580 ALD cycles, as-deposited). Figure~2 of the main text shows the smoothed initial calibration, in which the short spikes are removed; $R_{\mathrm{late}}$, all response values, and $t_{90}$ times were calculated from unsmoothed data. The first complete ascending program spans 0.05 to 5~\volpc{} \ce{H2} in \ce{N2} and is followed by the start of a second program. At each concentration, five 180~s exposures alternate with 180~s \ce{N2} recovery periods. Gray bands mark \ce{H2} exposure, and concentration labels are in \volpc. Asterisks mark the five initial conditioning exposures at 5~\volpc{} in each program. Low-concentration responses recur in the second program while the \ce{N2} baseline remains elevated.}
\label{fig:si-extended}
\end{figure}

The reference device provides the positive-response reference shown in Figure~2 of the main text. Its mean late responses at 0.2 and 3~\volpc{} \ce{H2} were 1.70 and 15.14~\%, respectively. The calibration curve flattens above approximately 3~\volpc. Linear fits give $R^2=0.985$ over 0.05 to 1~\volpc, $R^2=0.913$ over 1 to 5~\volpc{}, and $R^2=0.963$ over the full range. Similar concentration-dependent calibration regimes occur in closely spaced Pd nanoparticle arrays~\cite{xie_response_2012}. Baseline drift was 0.27~\% over 0.1 to 1~\volpc{} and 0.61~\% over 1 to 5~\volpc. A concentration-dependent drift component becomes apparent near 2~\volpc. Low-concentration responses recur at the start of a second, partially recorded ascending sequence while the \ce{N2} baseline remains elevated (Figure~\ref{fig:si-extended}). The baseline offset therefore persists into subsequent low-concentration cycling.

\begin{figure}[htbp]
\centering
\includegraphics[width=0.98\linewidth]{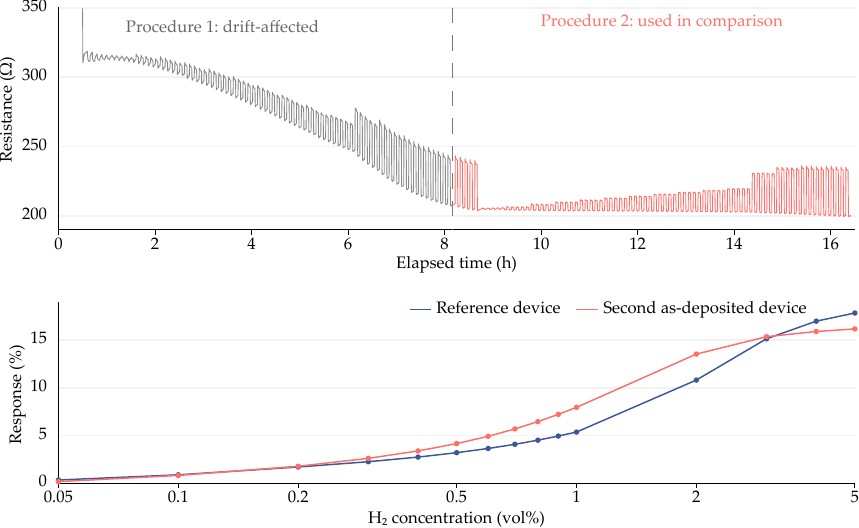}
\caption{Room-temperature comparison of the reference device and second as-deposited device, each prepared with 580 ALD cycles in separate runs using the same recipe. The upper panel shows the raw resistance of the second as-deposited device during two ascending concentration programs; the first run exhibits downward drift. The lower panel shows the late responses of the reference device (blue, first run) and second as-deposited device (coral, second run). Each cycle comprises 180~s of \ce{H2} exposure and 180~s of \ce{N2} recovery. Points show means, and bars show cycle standard deviations ($n=5$). The response is $100(R-R_0)/R_0$ averaged from 140 to 160~s after an onset determined from the resistance trace; $R_0$ is the median resistance from 40 to 10~s before onset.}
\label{fig:si-asdeposited}
\end{figure}

The second as-deposited device shows the same positive response trend after an initial run affected by drift (Figure~\ref{fig:si-asdeposited}). Under the common late-response definition, the two devices give $0.33\pm0.07$ and $0.16\pm0.08$~\% at 0.05~\volpc{}, and $17.85\pm0.01$ and $16.18\pm0.02$~\% at 5~\volpc. The resistance of the second as-deposited device is higher: 200 to 350~$\Omega$, compared with 39 to 46~$\Omega$ for the reference device. This difference may reflect reduced Pd deposition caused by temperature gradients during ALD, when heating was applied from below without an upper cover. The reference device responds more slowly at low concentrations than at high concentrations. An additional local maximum in response time occurs near 2~\volpc. Response and recovery are fastest at 5~\volpc.

\clearpage

\section{Temperature measurements and controls}
\label{si:temperature}

\subsection{Room-temperature comparison of two treated devices}
The treated device and replication device each received a single treatment for 10~min at 200~\degC{} in air and were measured near room temperature using the same program. The treated device provided one complete run, and the replication device provided two complete runs. The recordings for both devices had a median PT100 temperature of 23.70~\degC. This comparison was evaluated separately from the model-fitting data set. Gas switches were aligned to the first sustained change in the reference-sensor signal. The baseline window extended from 40 to 10~s before onset, and the late-response window extended from 140 to 160~s after onset; both were evaluated on a 1~s grid. Invalid startup telemetry was excluded before clock alignment.

\begin{figure}[htbp]
\centering
\includegraphics[width=.99\linewidth]{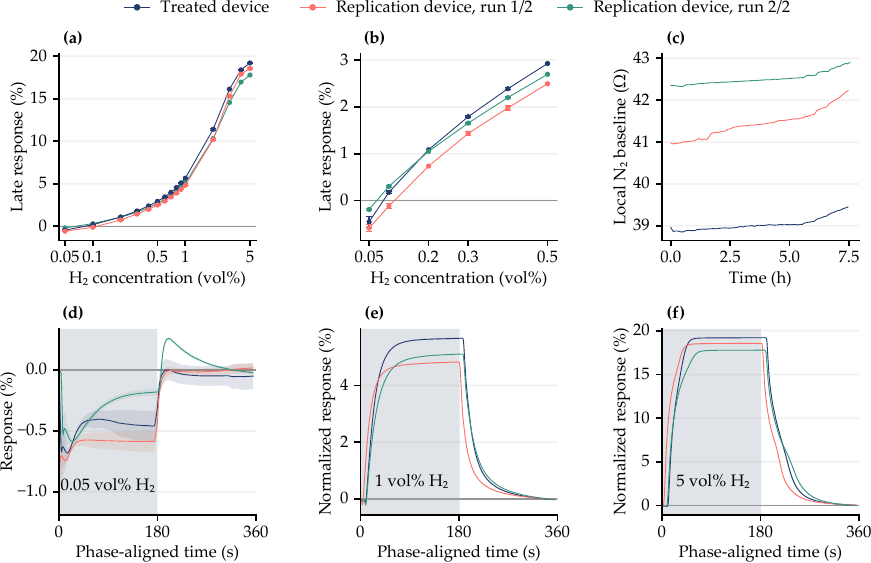}
\caption{Room-temperature responses of the treated device (blue) and replication device (two runs, coral and teal), both prepared with 580 ALD cycles and treated independently. The recordings for both devices had a median PT100 temperature of 23.70~\degC. (a,b) Mean late responses over the full and low-concentration ranges, respectively. (c) Local \ce{N2} baselines. (d--f) Mean normalized transients at 0.05, 1, and 5~\volpc{} \ce{H2}. Bars and bands show cycle standard deviations ($n=5$ per run). Exposure and recovery are aligned separately to the reference-gas transitions. Recovery begins at a nominal time of 180~s. Pale bands mark exposure.}
\label{fig:si-replication}
\end{figure}

At 5~\volpc{}, the late response was $19.20\pm0.04$~\% for the treated device and $18.56\pm0.09$ and $17.78\pm0.04$~\% for the two runs of the replication device. At 0.05~\volpc{}, the corresponding responses were $-0.45\pm0.12$, $-0.59\pm0.08$, and $-0.19\pm0.03$~\%, respectively. At 0.1~\volpc{}, the replication device gave $-0.12\pm0.06$ and $+0.30\pm0.02$~\%, placing this concentration at its null point. The low- and high-concentration polarity pattern therefore recurs across the two treated devices, although the response magnitude varies between devices and successive runs. Local \ce{N2} baselines were approximately 39~$\Omega$ for the treated device and 41 to 43~$\Omega$ for the replication device (Figure~\ref{fig:si-replication}).

\subsection{Thick-coating device at room temperature and 30~\degC}
\begin{figure}[htbp]
\centering
\includegraphics[width=.99\linewidth]{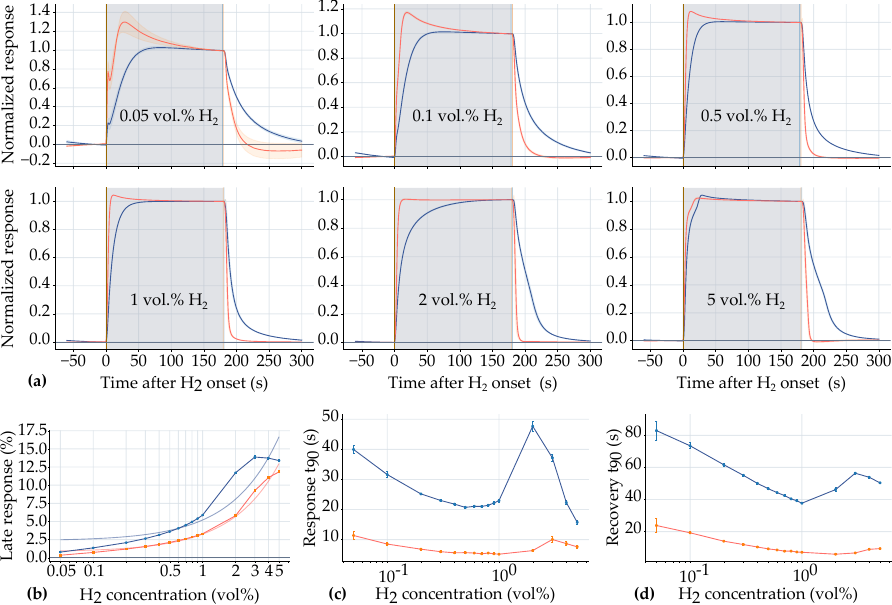}
\caption{Temperature-dependent response of the thick-coating device near room temperature (blue, median PT100 temperature 24.04~\degC) and at nominal 30~\degC{} (red, median 29.99~\degC). Each concentration includes five cycles, each comprising 180~s of \ce{H2} exposure and 180~s of \ce{N2} recovery. (a) Mean transients at 0.05, 0.1, 0.5, 1, 2, and 5~\volpc{}, normalized to the late exposure response. Bands show cycle standard deviations, and pale shading marks exposure. (b) Mean late response. (c,d) Mean response and recovery $t_{90}$, respectively. Bars in (b--d) show cycle standard deviations ($n=5$). Recovery is displayed from a nominal time of 180~s. The phase-transition and percolation-path annotations are schematic.}
\label{fig:si-thick}
\end{figure}

The thick-coating device, prepared with 1100 ALD cycles, was measured at median PT100 temperatures of 24.0 and 30.0~\degC{} (Figure~\ref{fig:si-thick}). At 30~\degC{}, the exposure transients overshoot at 0.05 and 0.1~\volpc{}, and the recovery transient undershoots after hydrogen removal at the lowest concentration. The room-temperature transients are slower and exhibit less overshoot. Heating reduces the exposure response and shortens both the response and recovery times. The pronounced room-temperature kinetic maximum near 2~\volpc{} is substantially reduced at 30~\degC. This trend is consistent with a contribution from phase conversion to the slower room-temperature response~\cite{yang2010joule}, together with faster hydrogen exchange and a shift in the channel balance.

\subsection{Post-treatment recording at 55~\degC}
\begin{figure}[htbp]
\centering
\includegraphics[width=0.9\linewidth]{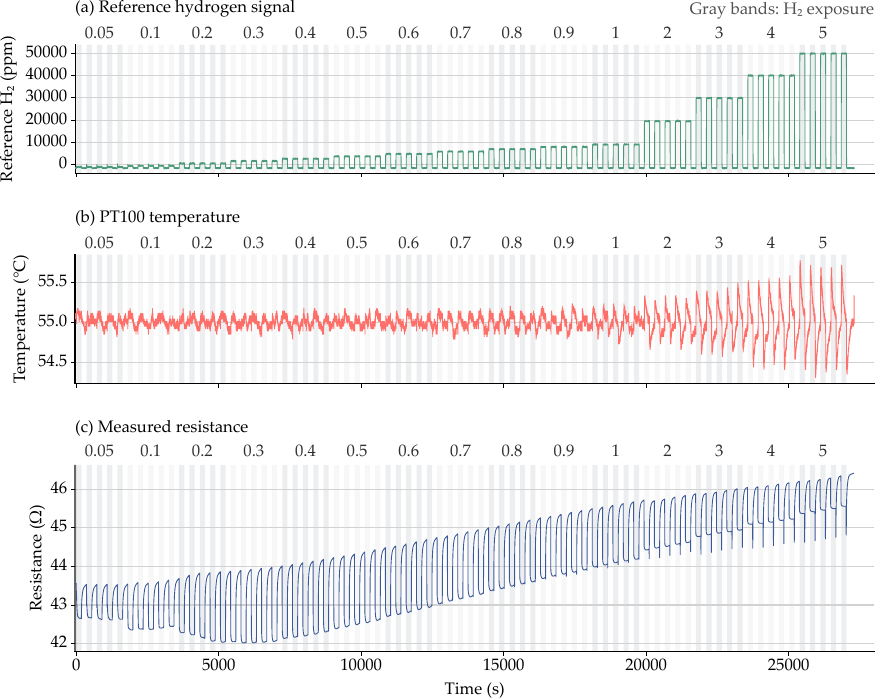}
\caption{Concentration program for the treated device at nominal 55~\degC{} after treatment in air at 200~\degC{} for 10~min. (a) Reference \ce{H2} signal. (b) PT100 temperature. (c) Raw resistance. All panels share an elapsed-time axis. Programmed concentrations range from 0.05 to 5~\volpc{}, with five 180~s exposures alternating with 180~s \ce{N2} recovery periods at each concentration. Gray bands mark exposure. The resistance decreases during every exposure.}
\label{fig:si-55C}
\end{figure}

Figure~\ref{fig:si-55C} shows the reference-gas, temperature, and resistance traces recorded simultaneously during the post-treatment program at 55~\degC. The local pre-exposure \ce{N2} baseline rises from 43.52 to 46.24~$\Omega$ over the ascending concentration program. From 100 to 168~s after onset, the cycle-averaged exposure response changes by less than 0.08 percentage points at every concentration. Over the corresponding recovery window, the cycle-averaged response continues to rise by 0.10 to 0.16 percentage points. Exposure and recovery thus evolve differently. The recovery behavior is consistent with continued relaxation of the device state; its possible contribution to the model mismatch at 55~\degC{} is discussed in Section~\ref{si:diagnostics}.

\section{Model specification and global fitting}
\label{si:model}

\subsection{Two electrical response channels}
The channel sum, reference resistance, and meanings of the scattering and connectivity contributions are defined in Equation~1 and the accompanying text of the main manuscript. The individual channel curves are model assignments, and their sum is compared with the measured response after the cycle-local normalization specified below.

\subsection{Hydrogen loading}
For a particle of diameter $D$, the equilibrium hydrogen loading $x_D$, expressed as H/Pd, follows a Lacher-type isotherm~\cite{lacher1937solubility},
\begin{equation}
\frac{1}{2}\ln\!\left(\frac{p_{\ce{H2}}}{p_0}\right)=\ln\!\left(\frac{x_D}{r-x_D}\right)-\frac{E_{\mathrm{eff}}x_D}{RT}+\frac{E_0(T,D)}{RT}.
\label{eq:si-lacher}
\end{equation}
$p_{\ce{H2}}=P_{\mathrm{tot}}c/100$, where $P_{\mathrm{tot}}=1$~bar and $c$ is the programmed concentration in \volpc. The reference pressure is $p_0=1$~bar, $r$ is the maximum site occupancy, and $E_{\mathrm{eff}}$ is the effective interaction between absorbed hydrogen atoms. Temperatures are expressed in kelvin, and energies are given per mole of H. The insertion energy comprises a size-independent term and a surface term,
\begin{equation}
E_0(T)=E_{0,\mathrm{ref}}-\frac{(T-T_{\mathrm{ref}})\,\Delta S_{\mathrm{f}}}{2},\qquad
E_0(T,D)=E_0(T)+\frac{u_{\mathrm{surf}}}{D},
\label{eq:si-e0}
\end{equation}
where $T_{\mathrm{ref}}=313.15$~K and $\Delta S_{\mathrm{f}}=-94.5$~J\,K$^{-1}$\,mol$^{-1}$~\ce{H2}. The factor of two converts the entropy to a per-mole-of-H basis. The ensemble loading is the number-weighted mean
\begin{equation}
x=\frac{\sum_i w_i x_{D_i}}{\sum_i w_i}
\label{eq:si-ensemble}
\end{equation}
over 15 equally spaced diameters spanning $\overline{D}\pm2.5\sigma_D$, subject to a lower limit of 12~nm, with Gaussian weights derived from the size distribution of the treated device. The electrical target laws are applied to this average loading. The imposed temperature dependence follows Pd--H thermodynamics. Particle size, surface contributions, and elastic coupling shift nanoparticle equilibria relative to bulk Pd~\cite{griessen2016thermodynamics}. Chemomechanical calculations also relate mechanical constraint to the miscibility gap~\cite{dyck_phase_2024}.

\subsection{Scattering channel}
\begin{equation}
S_{\mathrm{s}}^{\mathrm{eq}}(x,T)=A_{\mathrm{s}}(T)\left[\frac{\rho(x)}{\rho(0)}-1\right].
\label{eq:si-scatter}
\end{equation}
$\rho(x)/\rho(0)$ is the relative resistivity of PdH$_x$, interpolated from tabulated data~\cite{otterson1969absorption}. In the dilute $\alpha$ phase, resistivity increases approximately linearly with H/Pd~\cite{toth2005electrolytic}. $A_{\mathrm{s}}(T)$ scales the intrinsic PdH$_x$ response to that of the network. Thin-film measurements show that mechanical constraint affects solubility much more strongly than it affects resistivity at a given hydrogen content~\cite{wagner2010resistivity}. This observation supports treating hydrogen loading, $x(p_{\ce{H2}},T)$, separately from electrical transduction, $\rho(x)$.

\subsection{Connectivity channel}
\begin{equation}
S_{\mathrm{p}}^{\mathrm{eq}}(x)=-P_{\max}\left[1-\exp(-\kappa_{\mathrm{gap}}x)\right],\qquad
\kappa_{\mathrm{gap}}=\frac{10\,\eta_{\mathrm{H}}\overline{D}}{d^{\ast}}.
\label{eq:si-perc}
\end{equation}
$P_{\max}$ is the maximum fractional decrease in resistance. $\eta_{\mathrm{H}}$ is the linear lattice-expansion coefficient per H/Pd, $\overline{D}$ is the mean diameter in nanometers, and $d^{\ast}$ is an effective junction-distance scale in {\aa}ngstr{\"o}m. The factor of 10 converts nanometers to {\aa}ngstr{\"o}m. The exponential law represents a finite population of critical junctions that progressively close as hydrogen loading increases~\cite{kiefer_transition_2010,xie_response_2012}. The measured resistance constrains the combined sensitivity $\kappa_{\mathrm{gap}}$, from which $d^{\ast}$ is calculated using the fixed values of $\eta_{\mathrm{H}}$ and $\overline{D}$.

\subsection{Kinetics}
Each channel approaches its equilibrium target through first-order relaxation,
\begin{equation}
\frac{dS_j}{dt}=\frac{S_j^{\mathrm{eq}}-S_j}{\tau_j(T)},\qquad j\in\{\mathrm{s},\mathrm{p}\}.
\label{eq:si-kinetics}
\end{equation}
Both equilibrium targets are zero during \ce{N2} recovery. Each channel has its own time constant, which is the same for absorption and desorption within that channel. These effective time constants can incorporate surface exchange, hydrogen transport, lattice relaxation, and rearrangement of the electrical network. In shape-controlled Pd nanocrystals, hydrogen uptake and release exhibit different sensitivities to vertices and facets~\cite{johnson_facets_2019}, illustrating the microscopic processes represented by this effective description.

\subsection{Temperature parameterization}
$A_{\mathrm{s}}$ is parameterized by a bounded inverse-temperature relation,
\begin{equation}
a_{\mathrm{s}}(T)=a_{\mathrm{s,ref}}+\frac{E_{A_{\mathrm{s}}}}{R}\left(\frac{1}{T}-\frac{1}{T_{\mathrm{ref}}}\right),\qquad
A_{\mathrm{s}}(T)=A_{\min}+\frac{A_{\max}-A_{\min}}{1+\exp[-a_{\mathrm{s}}(T)]},
\label{eq:si-as}
\end{equation}
where $A_{\min}=0.02$, $A_{\max}=3$, and $a_{\mathrm{s,ref}}=\ln[(A_{\mathrm{s}}(T_{\mathrm{ref}})-A_{\min})/(A_{\max}-A_{\mathrm{s}}(T_{\mathrm{ref}}))]$. The scattering time constant is parameterized as
\begin{equation}
\ln\!\left[\frac{\tau_{\mathrm{s}}(T)}{\tau_{\mathrm{s}}(T_{\mathrm{ref}})}\right]=\frac{E_{\tau,\mathrm{s}}}{R}\left(\frac{1}{T}-\frac{1}{T_{\mathrm{ref}}}\right).
\label{eq:si-tau}
\end{equation}
$E_{A_{\mathrm{s}}}$ and $E_{\tau,\mathrm{s}}$ are the slopes of these imposed relations. The logistic bound means that $E_{A_{\mathrm{s}}}$ differs from the Arrhenius energy of $A_{\mathrm{s}}$ itself. The direct Arrhenius slope between the fitted values at 25 and 55~\degC{} is $R\ln(2.558/0.192)/(1/298.15\,\mathrm{K}-1/328.15\,\mathrm{K})=70$~kJ\,mol$^{-1}$.

\subsection{Global fitting procedure}
The joint fit used the post-treatment programs of the treated device at nominal temperatures of 25, 40, and 55~\degC. One complete run of 75 cycles was available at each of 25 and 55~\degC{}, and two complete runs of the same length, totaling 150 cycles, were available at 40~\degC. Each run comprised five cycles at 0.05 and 0.1~\volpc{}, at concentrations from 0.2 to 1.0~\volpc{} in steps of 0.1~\volpc{}, and at concentrations from 2 to 5~\volpc{} in steps of 1~\volpc. The five initial conditioning cycles at 5~\volpc{} initialized the model states but were excluded from the objective. Gas transitions were identified from the first sustained response of the reference sensor and represented as instantaneous concentration steps. The chamber time constant was fixed to zero. Each measured cycle was normalized to its local pre-exposure baseline $R_0$. For simulated cycle $c$, $b_{j,c}$ denotes the median channel state over the pre-exposure window from 40 to 10~s before onset. The quantities compared with the measurements are
\begin{equation}
\widetilde S_{j,c}(t)=\frac{S_j(t)-b_{j,c}}{1+b_{\mathrm{s},c}+b_{\mathrm{p},c}},\qquad
\widetilde S_c(t)=\widetilde S_{\mathrm{s},c}(t)+\widetilde S_{\mathrm{p},c}(t).
\label{eq:si-norm}
\end{equation}
Measured responses are expressed as $100(R-R_0)/R_0$, and simulated responses as $100\widetilde S_c$. The simulated states evolve continuously across gas switches and cycles, retaining the effects of incomplete recovery. Comparing responses relative to each cycle's local baseline prevents long-term baseline drift from dominating the fit.

The objective was a soft-$L_1$ loss minimized by constrained nonlinear least squares. Within each temperature, run, concentration, and gas-phase group, the five measured cycles were averaged on a common time grid and compared with the mean model prediction. Each difference was scaled by the larger of a 0.05 percentage-point noise floor and the group's median robust cycle-to-cycle standard deviation. A factor proportional to the inverse square root of the number of runs and time points balanced the groups. Consequently, each temperature, concentration, and gas phase had comparable influence despite the unequal numbers of runs. Penalty terms enforced the parameter constraints.

The size-distribution parameters were fixed at $\overline{D}=29.78$~nm and $\sigma_D=5.69$~nm. The remaining fixed quantities were $\eta_{\mathrm{H}}=0.063$, $r=0.6$, and $u_{\mathrm{surf}}=10$~kJ\,nm\,mol$^{-1}$~H. The parameter bounds were $-4$ to 8~kJ\,mol$^{-1}$~H for $E_{0,\mathrm{ref}}$, 10 to 30~kJ\,mol$^{-1}$~H for $E_{\mathrm{eff}}$, 0.02 to 3 for $A_{\mathrm{s}}$, 0 to 0.10 for $P_{\max}$, 1 to 300~(H/Pd)$^{-1}$ for $\kappa_{\mathrm{gap}}$, and 0.3 to 600~s for the time constants. The bounds for both temperature-slope parameters, $E_{A_{\mathrm{s}}}$ and $E_{\tau,\mathrm{s}}$, were $-200$ to 200~kJ\,mol$^{-1}$. Temperature dependence enters through $E_0(T)$, $A_{\mathrm{s}}(T)$, and $\tau_{\mathrm{s}}(T)$. The parameters $E_{\mathrm{eff}}$, $P_{\max}$, $\kappa_{\mathrm{gap}}$, and $\tau_{\mathrm{p}}$ are shared across temperatures. Nine parameters were fitted.

\subsection{Fitted parameters}
Table~\ref{tab:si-params} lists the point estimates. The size-independent insertion energy is $E_{0,\mathrm{ref}}=1.651$~kJ\,mol$^{-1}$~H at 40~\degC{}, giving $E_0(T)=0.942$, 1.651, and 2.360~kJ\,mol$^{-1}$~H at 25, 40, and 55~\degC{}, respectively. At the mean diameter, the surface term $u_{\mathrm{surf}}/D$ adds 0.336~kJ\,mol$^{-1}$~H. The scattering amplitude decreases successively from 2.558 to 1.055 and 0.192, an overall decrease by a factor of 13.3. The scattering time constant decreases successively from 25.78 to 11.94 and 5.94~s, an overall decrease by a factor of 4.3, with $E_{\tau,\mathrm{s}}=39.82$~kJ\,mol$^{-1}$. This energy lies within the range of apparent activation energies reported for hydrogen absorption and desorption in Pd nanoparticles~\cite{stolas2020surfactants,alekseeva2021grain}. Heated Pd nanowires likewise exhibit smaller response magnitudes and faster kinetics~\cite{yang2010joule}. The shared connectivity parameters are $P_{\max}=0.03704$, $\kappa_{\mathrm{gap}}=165.50$~(H/Pd)$^{-1}$, and $\tau_{\mathrm{p}}=17.77$~s. With the fixed expansion coefficient and mean diameter, $\kappa_{\mathrm{gap}}$ corresponds to $d^{\ast}=0.1134$~{\AA}.

\begin{table}[htbp]
\centering
\caption{Point estimates from the joint fit. $E_0(T)$ is the size-independent insertion energy; the isotherm includes the additional term $u_{\mathrm{surf}}/D$ for each particle diameter. The null point is the modeled equilibrium concentration at which the total response changes sign. The null point at 55~\degC{} is extrapolated beyond the tested concentration range.}
\label{tab:si-params}
\small
\begin{tabular}{cccccc}
\toprule
$T$ & $E_0(T)$ & $A_{\mathrm{s}}(T)$ & $\tau_{\mathrm{s}}(T)$ & $\tau_{\mathrm{p}}$ & Null point \\
(\degC) & (kJ\,mol$^{-1}$~H) & & (s) & (s) & (\volpc{} \ce{H2}) \\
\midrule
25 & 0.942 & 2.558 & 25.78 & 17.77 & 0.120 \\
40 & 1.651 & 1.055 & 11.94 & 17.77 & 1.361 \\
55 & 2.360 & 0.192 & 5.94  & 17.77 & 25.90 \\
\bottomrule
\end{tabular}
\end{table}

\subsection{Parameter sensitivity and identifiability}
Replacing the size distribution of the treated device with the 1100-cycle distribution changes the total response trajectories by at most $3.2\times10^{-5}$ percentage points over 25\,800 modeled points, while shifting the fitted insertion energy and the size-derived value of $d^{\ast}$. The total response is therefore well constrained, whereas the assignment of morphology-dependent parameters is not. The interaction parameter $E_{\mathrm{eff}}$ reaches its lower bound of 10~kJ\,mol$^{-1}$~H. In the Lacher model, the critical temperature is $T_{\mathrm{c}}=rE_{\mathrm{eff}}/(4R)$, giving 180~K at this bound and 541~K for the bulk value $E_{\mathrm{eff}}\approx30$~kJ\,mol$^{-1}$~H. The fitted branch is consequently single-valued at all three temperatures. The model includes no explicit $\alpha$--$\beta$ phase fraction, transformation kinetics, or hysteresis. Any room-temperature contribution from phase conversion would be reflected in the effective parameters and structured residuals. The local Jacobian condition number is 1973. The strongest correlations are $r=0.967$ between $E_{0,\mathrm{ref}}$ and $E_{A_{\mathrm{s}}}$, and $r=-0.956$ between the scattering reference time constant and its temperature slope.

\section{Full model comparisons and residual diagnostics}
\label{si:diagnostics}

\begin{figure}[htbp]
\centering
\includegraphics[width=.87\linewidth]{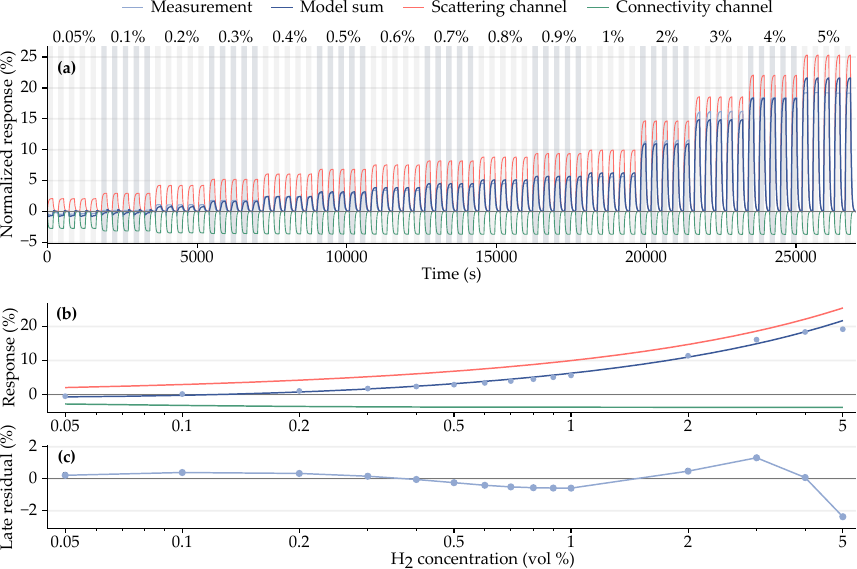}
\caption{Joint two-channel model at nominal 25~\degC{} (median PT100 temperature 23.70~\degC) for the treated device. (a) First run displayed on a concatenated timeline, with five cycles per concentration, each comprising 180~s of \ce{H2} exposure and 180~s of \ce{N2} recovery. Gray bands mark exposure. The measured response (pale blue), model sum (blue), scattering contribution (red), and connectivity contribution (green) are each referenced to the cycle-local baseline. (b) Equilibrium channel responses relative to zero loading (lines) and measured late responses (points). (c) Mean late residuals, defined as measurement minus model, from 140 to 160~s on the common 4~s grid. Bars in (b,c) show cycle standard deviations ($n=5$).}
\label{fig:si-model25}
\end{figure}

\begin{figure}[htbp]
\centering
\includegraphics[width=.87\linewidth]{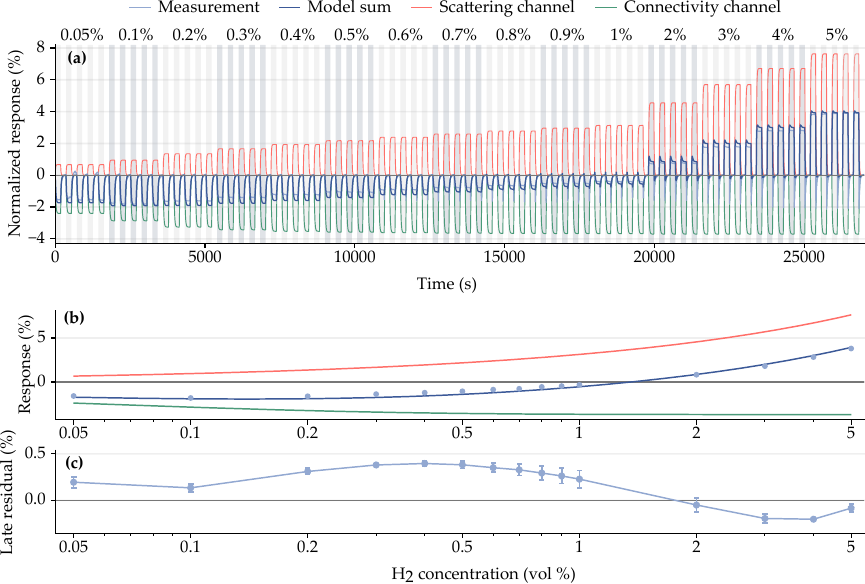}
\caption{Joint two-channel model at 40~\degC{} for the treated device. (a) First of two runs displayed on a concatenated timeline. Gray bands mark \ce{H2} exposure. The measured response (pale blue), model sum (blue), scattering contribution (red), and connectivity contribution (green) are referenced to the cycle-local baseline. (b) Equilibrium channel responses relative to zero loading (lines) and measured late responses (points). (c) Mean late residuals, defined as measurement minus model. Points and bars in (b,c) show means and cycle standard deviations across both runs ($n=10$).}
\label{fig:si-model40}
\end{figure}

\begin{figure}[htbp]
\centering
\includegraphics[width=.87\linewidth]{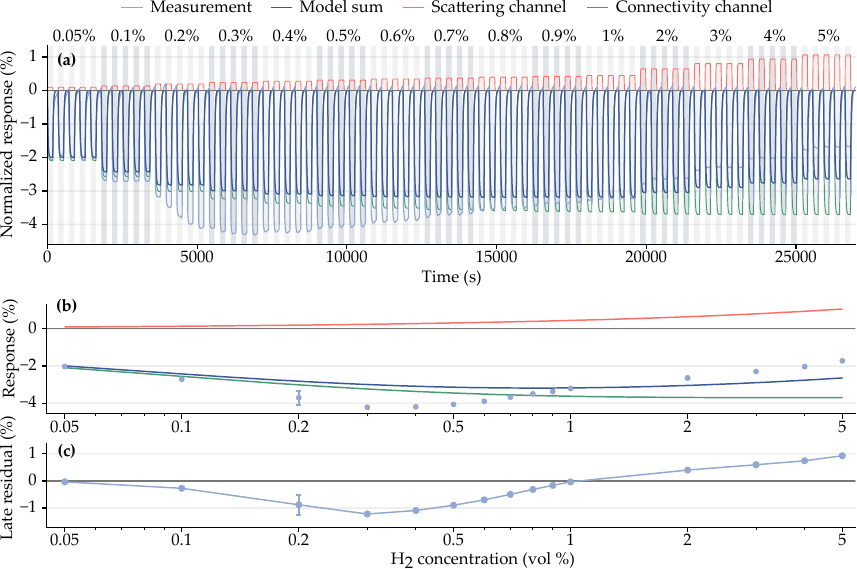}
\caption{Joint two-channel model at 55~\degC{} for the treated device. (a) First run displayed on a concatenated timeline. Gray bands mark \ce{H2} exposure. The measured response (pale blue), model sum (blue), scattering contribution (red), and connectivity contribution (green) are referenced to the cycle-local baseline. (b) Equilibrium channel responses relative to zero loading (lines) and measured late responses (points). (c) Mean late residuals, defined as measurement minus model. Bars in (b,c) show cycle standard deviations ($n=5$).}
\label{fig:si-model55}
\end{figure}

Figures~\ref{fig:si-model25} to \ref{fig:si-model55} show the complete first run at each temperature, together with the modeled equilibrium response and late-response residuals. Experimental points are averages of the raw resistance samples from 140 to 160~s. The modeled late-response curves are averages of predictions stored at 4~s intervals over the same window. Full-run plots concatenate nominal 360~s cycles, whereas the state calculations retain the recorded gas-switch chronology.

\clearpage
\subsection{Residual statistics}
\begin{figure}[htbp]
\centering
\includegraphics[width=0.88\linewidth]{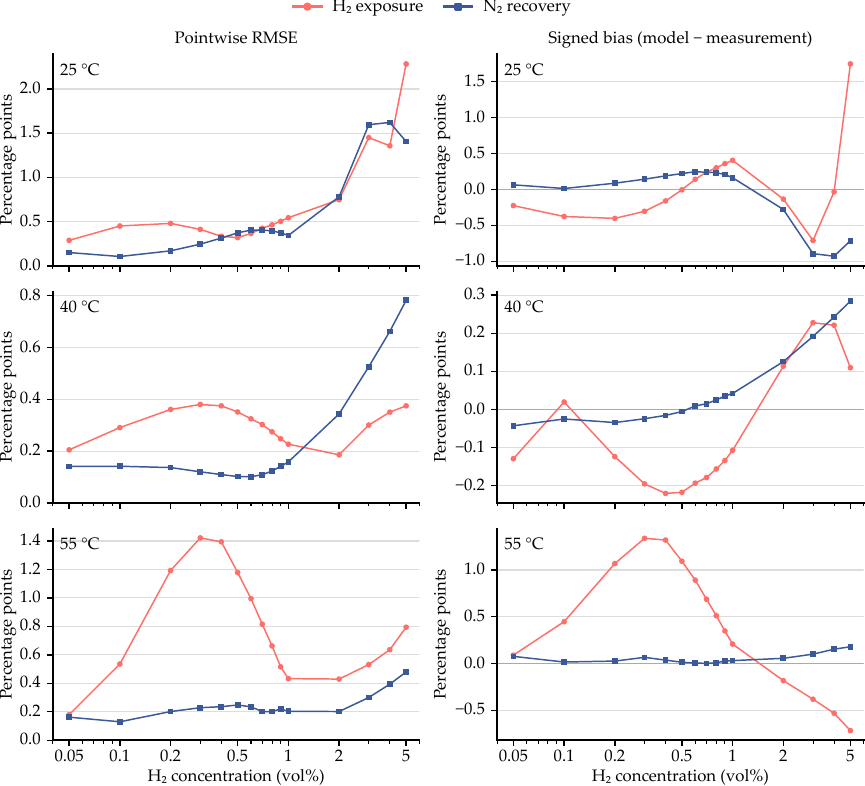}
\caption{Concentration-resolved diagnostics of the joint fit to the treated device at nominal 25, 40, and 55~\degC{} (top to bottom). Left panels show pointwise RMSE; right panels show signed bias, defined as model minus measurement. Red circles represent \ce{H2} exposure, and blue squares represent \ce{N2} recovery. Values are expressed in percentage points of $100(R-R_0)/R_0$. Each statistic includes five cycles per concentration and run. The statistics at 40~\degC{} are averaged across both runs.}
\label{fig:si-residuals}
\end{figure}

The temperature-balanced mean group RMSE is 0.474 percentage points. The overall pointwise RMSE and mean absolute error are 0.569 and 0.314 percentage points, respectively. The pointwise RMSEs at 25, 40, and 55~\degC{} are 0.828, 0.321, and 0.636 percentage points, respectively. At 25~\degC{} and 5~\volpc{}, the exposure RMSE is 2.28 percentage points, with a model-minus-measurement bias of 1.75 percentage points. Figure~\ref{fig:si-residuals} shows these deviations by concentration and gas phase.

At 55~\degC{}, the measured late response reaches a minimum of $-4.22$~\% at 0.3~\volpc{}, whereas the fitted minimum is $-3.18$~\% near 0.8~\volpc. At 0.5~\volpc{}, the measured response is $-4.04$~\%, and the fitted scattering and connectivity contributions are $+0.31$ and $-3.45$~\%, respectively, giving a total of $-3.14$~\%. The shared connectivity limit of $100P_{\max}=3.704$~\% restricts the depth of the modeled minimum, and cycle-local normalization does not remove this restriction. The evolution of the baseline and recovery at 55~\degC{} (Figure~\ref{fig:si-55C}) suggests that the junction weighting may change during the run.

\clearpage

\section{Cycle-bootstrap uncertainty and temperature holdout tests}
\label{si:bootstrap}

Complete cycles, each comprising paired exposure and recovery phases, were resampled with replacement within each temperature, run, and concentration group. Resampling preserved five cycles per group and the temporal structure within each cycle. The gas program, morphology, temperature assignments, and model structure were held fixed. Each constrained soft-$L_1$ refit began near the global optimum, used the original bounds, and allowed up to 500 function evaluations. All 200 refits converged. Table~\ref{tab:si-boot} reports intervals from the 2.5th to the 97.5th percentile for the nine fitted and six derived quantities. These intervals quantify cycle-to-cycle variation under within-group exchangeability. The interval for $E_{\mathrm{eff}}$ is affected by its active lower bound.

\begin{table}[htbp]
\centering
\caption{Conditional 95~\% cycle-bootstrap percentile intervals from 200 refits, shown alongside the point estimates from the all-temperature fit.}
\label{tab:si-boot}
\small
\begin{tabular}{llrc}
\toprule
Quantity & Units & Point estimate & Percentile interval \\
\midrule
$E_{0,\mathrm{ref}}$ & kJ\,mol$^{-1}$~H & 1.651 & [1.639, 1.716] \\
$E_{\mathrm{eff}}$ (lower bound) & kJ\,mol$^{-1}$~H & 10.000 & [10.000, 10.118] \\
$A_{\mathrm{s}}(40~\degC)$ & -- & 1.055 & [1.045, 1.082] \\
$E_{A_{\mathrm{s}}}$ & kJ\,mol$^{-1}$ & 123.111 & [121.431, 130.383] \\
$\tau_{\mathrm{s}}(40~\degC)$ & s & 11.942 & [11.604, 12.076] \\
$E_{\tau,\mathrm{s}}$ & kJ\,mol$^{-1}$ & 39.817 & [39.269, 40.965] \\
$100P_{\max}$ & \% & 3.704 & [3.666, 3.719] \\
$\kappa_{\mathrm{gap}}$ & (H/Pd)$^{-1}$ & 165.497 & [159.908, 178.421] \\
$\tau_{\mathrm{p}}$ & s & 17.774 & [17.355, 17.966] \\
\midrule
$A_{\mathrm{s}}(25~\degC)$ & -- & 2.558 & [2.541, 2.621] \\
$A_{\mathrm{s}}(55~\degC)$ & -- & 0.192 & [0.178, 0.195] \\
$\tau_{\mathrm{s}}(25~\degC)$ & s & 25.776 & [25.302, 25.980] \\
$\tau_{\mathrm{s}}(55~\degC)$ & s & 5.936 & [5.669, 6.056] \\
Null point 25~\degC & \volpc{} \ce{H2} & 0.120 & [0.115, 0.123] \\
Null point 40~\degC & \volpc{} \ce{H2} & 1.361 & [1.350, 1.364] \\
\bottomrule
\end{tabular}
\end{table}

As a sensitivity check, one temperature program was omitted at a time. The model was refitted to the other two programs using six optimization starts, and the omitted program was predicted using its recorded gas-switch times and the same baseline convention. All three fits converged. Table~\ref{tab:si-holdout} reports the pointwise RMSE. The holdout test at 40~\degC{} involves interpolation between the training temperatures; the other two tests involve extrapolation. The refit that omitted 40~\degC{} reached the 200~kJ\,mol$^{-1}$ upper bound of $E_{A_{\mathrm{s}}}$. The refit that omitted 25~\degC{} reached the 300~(H/Pd)$^{-1}$ upper bound of $\kappa_{\mathrm{gap}}$. The fits to pairs of temperatures did not predict the response at the omitted temperature. The holdout errors therefore quantify the limits of extrapolation rather than a failure of the joint description.

\begin{table}[htbp]
\centering
\caption{Prediction errors from temperature holdout tests, expressed in percentage points. The all-temperature fit provides the descriptive benchmark.}
\label{tab:si-holdout}
\small
\begin{tabular}{cccc}
\toprule
Held-out $T$ (\degC) & Training $T$ (\degC) & All-temperature RMSE & Held-out RMSE \\
\midrule
25 & 40, 55 & 0.828 & 4.257 \\
40 & 25, 55 & 0.321 & 2.662 \\
55 & 25, 40 & 0.636 & 0.834 \\
\bottomrule
\end{tabular}
\end{table}

\section{Measurement setup and data processing}
\label{si:setup}

\subsection{Gas mixing and electrical measurements}
\begin{figure}[htbp]
\centering
\includegraphics[width=0.6\linewidth]{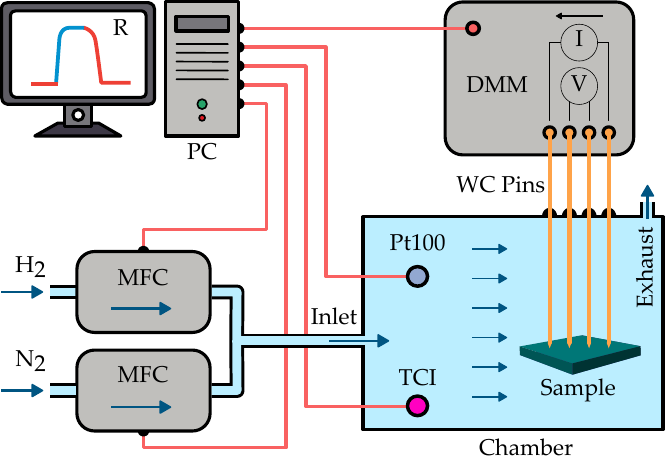}
\caption{Four-point resistance measurement during controlled hydrogen exposure. Two mass-flow controllers (MFC) mix 5~\volpc{} \ce{H2} in \ce{N2} with pure \ce{N2} at a total flow rate of 1 standard liter per minute through a 7.05~mL chamber. Four tungsten-carbide (WC) pins contact the sample for voltage measurement with an alternating $\pm1$~mA current. A PT100 measures the temperature at a sample corner, and a thermal-conductivity hydrogen sensor (TCI) provides a reference gas signal. A computer records all signals.}
\label{fig:si-setup}
\end{figure}

Figure~\ref{fig:si-setup} shows the measurement chamber. Two mass-flow controllers (MC series, Alicat Scientific) mix a premixed stream of 5~\volpc{} \ce{H2} in \ce{N2} with pure \ce{N2} at a total flow rate of 1 standard liter per minute. The chamber has a volume of 7054~mm$^3$ and operates near 1~bar. Using the nominal flow rate at chamber conditions gives a residence time $V/Q$ of 0.42~s and an ideal well-mixed exchange time $t_{90}=\ln(10)\,V/Q$ of 0.98~s. The actual volumetric flow rate depends on the controller reference conditions and the chamber temperature and pressure.

Four tungsten-carbide pins contact the sample. A Keithley 6220 supplies alternating currents of $+1$ and $-1$~mA, and a Keithley 2182 nanovoltmeter measures the voltage. A Raspberry Pi Pico 2 H records the resistance, pressure, temperature, and reference \ce{H2} signal. The sensing element of an Infineon Hydrogen 2 Click module (TCIXTMA1) provides the reference signal, and a Bosch BMP581 monitors the chamber pressure. A four-wire Class A PT100 in contact with a sample corner measures the temperature through a MAX31865 converter. This measurement provides feedback to a Kapton heater powered at 5~V through a MOSFET. A host computer logs the data through USB. The reference and pressure signals are used only for monitoring and phase alignment; no amplitude correction is applied to the resistance.

\subsection{Gas-phase alignment}
The reference device and the second as-deposited device were measured without synchronized reference-gas or temperature telemetry. For these recordings, the programmed switches were aligned to the largest absolute local derivative of an 11-point-smoothed resistance trace within a bounded window around each expected switch. The detected \ce{H2} exposure intervals lasted 160 to 205~s. The first run of the second as-deposited device was excluded because strong downward drift overlapped the ascending concentration program.

In the paired aging comparison, the reference-sensor transitions lagged the resistance changes in the aged room-temperature and 40~\degC{} recordings. These recordings were aligned to the largest absolute derivative of a seven-point-smoothed resistance trace near each programmed switch, giving exposure intervals of 167 to 188~s. All other recordings used the reference-sensor timing. Each complete run began with five conditioning cycles at 5~\volpc{}, followed by five cycles at each of 15 concentrations. The aged series included one complete run at each of room temperature and 40~\degC{}, and two runs at 55~\degC. The treated series included one run at each of nominal 25 and 55~\degC{}, and two runs at 40~\degC. The 40~\degC{} recording of the thick-coating device was excluded because a resistance discontinuity occurred from approximately 3~\volpc{} onward.

\subsection{Response definitions}
$R_0$ is the median resistance from 40 to 10~s before \ce{H2} onset. The late response is the mean of $100(R-R_0)/R_0$ from 140 to 160~s after onset. This finite-time definition applies to both positive and negative transients. The response time $t_{90}$ is the first crossing of 90~\% of the signed change from $R_0$ to the late level. The recovery time is the first crossing of 90~\% of the signed change from the resistance at switch-off to the mean resistance from 140 to 160~s after the switch to \ce{N2}. Each search is restricted to the corresponding gas phase. Because these definitions use the first crossing, the reported times can include early crossings during overshoot. Programmed concentrations and gas-phase times are used without an independent correction for gas delivery.

\end{document}